\documentclass{JHEP3}
\usepackage{cite}
\usepackage[english]{babel}
\usepackage{amsmath, amsthm, amssymb, amsfonts}
\usepackage{multicol}
\usepackage{verbatim}
\usepackage{bbold}
\usepackage{bbm}
\usepackage{mathbbol}
\usepackage{eucal}
\newcommand{\cA}{{\cal A}}  \newcommand{\cB}{{\cal B}}
\newcommand{\cC}{{\cal C}}  \newcommand{\cD}{{\cal D}}
  
\newcommand{\cG}{{\cal G}}  
  
  \newcommand{\cL}{{\cal L}}
\newcommand{\cM}{{\cal M}}  \newcommand{\cN}{{\cal N}}
\newcommand{\cO}{{\cal O}}

  \newcommand{\cV}{{\cal V}}

\newcommand{\be}{\begin{equation}} \newcommand{\ee}{\end{equation}}
\newcommand{\bea}{\begin{eqnarray}} \newcommand{\eea}{\end{eqnarray}}
\newcommand{\beann}{\begin{eqnarray*}}  \newcommand{\eeann}{\end{eqnarray*}}
\newcommand{\bfig}{\begin{figure}} \newcommand{\efig}{\end{figure}}
\newcommand{\ba}{\begin{array}} \newcommand{\ea}{\end{array}} 
\newcommand{\bcen}{\begin{center}} \newcommand{\ecen}{\end{center}}
\newcommand{\btab}{\begin{tabular}} \newcommand{\etab}{\end{tabular}}

\newcommand{\Eq}[1]{(\ref{#1})}
\newcommand{\vev}[1]{\left\langle{#1}\right\rangle}

\newcommand{\ra}{\rightarrow}

\newcommand{\e}{{\rm e}}

\newtheorem{Proposition}{Proposition}[section]

\newtheorem{Theorem}{Theorem}[section]
\newtheorem{Lemma}{Lemma}[section]
\newtheorem{Corrolary}{Corrolary}[section]

\newcommand{\bp}{\begin{Proposition}}   \newcommand{\ep}{\end{Proposition}}
\newcommand{\bt}{\begin{Theorem}}   \newcommand{\et}{\end{Theorem}}
\newcommand{\bl}{\begin{Lemma}}     \newcommand{\el}{\end{Lemma}}
\newcommand{\bc}{\begin{Corrolary}} \newcommand{\ec}{\end{Corrolary}}

\def\e{\eta}
\def\ep{\varepsilon}

\def\w{\omega}

\def\pd{\partial}

\title{Ward identities and relations between conductivities and viscosities in holography}
\author{Carlos Hoyos, David Rodr\'{\i}guez Fern\'andez,\\
\footnotesize{Department of Physics, Universidad de Oviedo,  Avda. Calvo Sotelo 18, 33007, Oviedo, Spain.}\\
E-mail: \email{hoyoscarlos@uniovi.es}, \email{rodriguezferdavid@uniovi.es}}
\preprint{FPAUO-15/14}
\abstract{We derive relations between viscosities and momentum conductivity in $2+1$ dimensions by finding a generalization of holographic Ward identities for the energy-momentum tensor. The generalization is novel in the sense that it goes beyond the usual identities obtained from holographic renormalization. Our results are consistent with previous field theory analysis. The main tools we use are a constant `probability current' in the gravity dual, that we are able to define for any system of linear ODEs, and parity symmetry. 
}

\begin{document}


\section{Introduction}

Holography, in the sense of the AdS/CFT correspondence  \cite{Maldacena:1997re,Gubser:1998bc,Witten:1998qj} and its generalizations, has been used as a tool to study strongly coupled systems that are otherwise intractable or notoriously difficult to deal with. An important aspect has been the derivation of fluid properties, in particular those associated to the transport of conserved currents such as energy, momentum or charge. Transport is characterized by transport coefficients, that can be defined from correlators of conserved currents through Kubo formulas, or in other ways such as constitutive relations in hydrodynamics. In holography they have been computed both ways, following the seminal works \cite{Policastro:2001yc,Bhattacharyya:2008jc}.  An alternative way to identify the coefficients is through the derivative expansion of the equilibrium partition function as in \cite{Banerjee:2012iz,Jensen:2012jh}, although dissipative coefficients are not captured with this method. One should bear in mind that those definitions are not always equivalent, when we discuss transport coefficients we will be referring to those derived from correlators.

Not all the transport coefficients one can possibly define are independent. For correlators of conserved currents there are Ward identities that impose relations among them, thus constraining some of the transport coefficients. In some cases these relations lead to interesting effects, an example is the relation between Hall conductivity and Hall viscosity found in \cite{Hoyos:2011ez} for Quantum Hall systems. Both coefficients are interesting from the point of view of the characterization of topological phases. For instance, in a Quantum Hall system the Hall conductivity is proportional to the filling fraction while the Hall viscosity depends on the shift \cite{PhysRevLett.69.953,PhysRevB.79.045308,PhysRevB.84.085316}, both of which take discrete values and remain fixed under small deformations. However, in order to determine the viscosity  one in principle needs to deform the material and measure the resulting stress, while the conductivity can be determined by a much simpler measurement of an electric current. This situation is helped by the relation between the two. In the presence of an inhomogeneous electric field, the Hall current receives a correction, which to leading order in derivatives is
\begin{equation}
J^i_H\simeq  \sigma_H^{(0)} E^i+\sigma_H^{(2)} \nabla^2 E^i
\end{equation}
The coefficient $\sigma_H^{(2)}$ depends on the Hall viscosity and other quantities that can be determined independently. This in principle allows to measure the Hall viscosity via inhomogeneous electric fields, which may be easier to realize experimentally than a direct measurement of the viscosity. Originally the relation was obtained form an effective action approach, but later it was shown for Galilean invariant systems that the relation can be derived from Ward identities \cite{Bradlyn:2012ea}. 

These relations were further generalized for any system with rotational invariance in \cite{Hoyos:2015yna}. For the general case (not Galilean invariant), the transport coefficients that are directly related are conductivities in the momentum current (or thermal conductivities) and viscosities. Charge conductivities also enter when the magnetic field is nonzero. In terms of retarded correlators of the energy-momentum tensor
\begin{equation}
\Gamma^{\mu\nu\alpha\beta}(x,\hat{x})=\vev{T^{\mu\nu}(x)T^{\alpha\beta}(\hat{x})}_R,
\end{equation}
the relevant Ward identity is, in the absence of external sources  ($i,j=1,2$ label the spatial directions)
\begin{equation}
\partial_0 \hat{\partial}_0 \Gamma^{0 i 0 j}(x,\hat{x})+\partial_k \hat{\partial}_l \Gamma^{k i l j}(x,\hat{x})\simeq 0.
\end{equation}
The right hand side might contain contact terms but otherwise it's zero because of the conservation of the energy-momentum tensor.  The identity can be derived combining the two identities
\begin{equation}\label{wardids}
\partial_\mu \Gamma^{\mu\nu\alpha\beta}(x,\hat{x})\simeq 0,\ \ \hat{\partial}_\alpha \Gamma^{\mu\nu\alpha\beta}(x,\hat{x})\simeq 0.
\end{equation}
For readers familiar with the AdS/CFT correspondence it might seem that these identities have been derived already from the holographic renormalization procedure \cite{deHaro:2000xn,Papadimitriou:2004rz}, in particular the Ward identities for charged $2+1$ dimensional systems were studied in some detail in \cite{Hartnoll:2007ip,Herzog:2009xv,Lindgren:2015lia}. This is partially true, there is a set of Ward identities that hold for the correlators of the energy-momentum tensor with any other operators $\cO_1,\dots,\cO_n$ 
\begin{equation}
\partial_\mu \vev{ T^{\mu\nu}(x) \cO_1(x_1)\cdots \cO_n(x_n)}=0.
\end{equation}
In the case of asymptotically AdS spacetimes\footnote{Generalizations for other geometries have been discussed in \cite{Kanitscheider:2008kd,Papadimitriou:2011qb,Gouteraux:2011qh,Caldarelli:2013aaa,Chemissany:2014xsa,Hartong:2014oma}.} this follows from `kinematics', it is not necessary to know the full geometry but the identity follows from the asymptotic expansion and the equations of motion. However, the second identity in \eqref{wardids} does not follow directly from the asymptotic expansion, it requires further input. 

So far the relations between viscosities and conductivities in holography could only be checked by direct computation of the correlators, but a general `kinematic' argument should exist, since they follow from symmetries and will be valid in any field theory. In this paper we make a first step towards generalizing Ward identities for two-point functions of the energy-momentum in $2+1$ dimensions. We will establish the relation between the parity even components of the conductivities and the shear and bulk viscosities, our argument relies on constructing a quantity which is independent of the radial direction in the bulk geometry and taking advantage of parity symmetry. The radially independent quantity can be seen as a ``probability current'' for the solutions to the linear equations of motion.  We give a general prescription on how to construct the probability current for any linear system of second order ordinary differential equations and apply it to a specific set of theories consisting of $3+1$-dimensional gravity coupled to a scalar field. In the context of scattering in black hole geometries, the probability current is the flux through a surface at a fixed value of the radial coordinate, in particular the flux through the horizon, so it determines the absorption by the black hole. In simple cases, such as a probe scalar field, one can see that the probability current is proportional to the spectral function of the dual operator. In more general cases it is still a combination of correlators, but we do not have a clean interpretation for it, we will use it as a mathematical device to derive the Ward identities.

The paper is organized as follows: In \S~\ref{shearid} we derive the identities that relate shear viscosity and momentum conductivity in a conformal field theory (CFT). In \S~\ref{bulkid} we generalize our construction to theories with explicit breaking of conformal invariance via a relevant deformation and derive identities that relate bulk viscosity and momentum conductivity. In \S\S~\ref{probcurr} we present the general construction of the probability current. We discuss the results in \S~\ref{discuss} and a possible application of the probability current to compute the spectrum of normalizable modes. We have gathered a collection of technical results in the Appendices at the end of the paper.

\section{Shear identity in a CFT}\label{shearid}

We will start by working out a simple example that will serve to illustrate the procedure we are proposing to derive generalized Ward identities, without the technical complications of more involved cases. 

The simplest identity we can check using holography is the relation between shear viscosity and thermal conductivity in a CFT. We will restrict the analysis to $(2+1)$-dimensions, although it can be generalized to any number of dimensions. We will assume that the CFT has a gravity dual and that quantum and higher derivative corrections on the gravity side are small,\footnote{On the field theory side this means a large-$N$ and strong coupling approximation.} so it can be well approximated by classical Einstein gravity coupled to matter. For simplicity we will consider states  where the CFT is at finite temperature but there are no other sources of breaking of conformal invariance (explicit or spontaneous). This implies that the effect of matter is simply to introduce a negative cosmological constant $\Lambda= -3/L^2$
\begin{equation}\label{classac}
S=\frac{1}{16\pi G_N}\int d^4 x\, \sqrt{-g}\left(R-2\Lambda\right),
\end{equation}
where $G_N$ is the four-dimensional Newton's constant. The geometry dual to the thermal state of a CFT is the $AdS_4$ black brane 
\begin{equation}
ds^2=\frac{L^2}{z^2}\left(-f(z) dx_0^2+dx_1^2+dx_2^2+\frac{dz^2}{f(z)}\right), \ \ f(z)=1-\frac{z^3}{z_H^3}.
\end{equation}
Where $L$ is the $AdS$ radius. The conformal $AdS$ boundary is at $z=0$, while $z=z_H$ is the position of the black brane horizon. We can set $z_H=1$  by rescaling the coordinates
\begin{equation}
z\to z_H z,\ \ x^\mu\to z_H x^\mu.
\end{equation}
All dimensionful quantities will be given in units of $z_H$. Physical units can be restored by introducing $z_H$ factors using dimensional analysis and then replacing the $z_H$ dependence by a dependence on the temperature of the black brane
\be \label{hawtemp}
T=\frac{3}{4\pi z_H}\,.
\ee

Two-point retarded correlation functions of the energy-momentum tensor can be computed using AdS/CFT by solving for linearized fluctuations of the metric around the black brane background and imposing ingoing boundary conditions at the horizon \cite{Son:2002sd}. We will work in the radial gauge where $\delta g_{Mz}=0$\footnote{We will employ capital latin indexes for the bulk coordinates, greek indexes for the boundary coordinates. Latin lower case indexes will run for the spatial $x_1,x_2$ components.} and we will expand in plane waves with momentum along the $x_1$ direction 
\begin{equation} \label{fluctuation1}
\delta g_{\mu\nu}=-\frac{L^2}{z^2}\int \frac{d^3 k}{(2\pi)^3} e^{i k_\mu x^\mu} h_{\mu\nu}(z).
\end{equation}
For the calculation of the shear viscosity we only need to turn on the $h_{12}$ and $h_{02}$ components, and in the calculations we will fix the momentum to be $k^\mu=(\omega,k,0)$ without loss of generality.

Varying the action \eqref{classac} with respect to the metric $g_{MN}$ yields Einstein's equations 
\begin{equation}
R_{MN}-\frac{1}{2}g_{MN} R -\frac{3}{L^2} g_{MN}=0.
\end{equation}
Expanding to linear order in the fluctuations, we find three equations, two dynamical and one constraint 
\begin{equation} \label{eoms}
\begin{split}
0=& h_{12}''-\left( \frac{2}{z}-\frac{f'}{f}\right)h_{12}'+\frac{\omega^2}{f^2}h_{12}+\frac{\omega k}{f^2}h_{02}, \\
0=& h_{02}''-\frac{2}{z}h_{02}'-\frac{k^2}{f}h_{02}-\frac{\omega k}{f}h_{12}, \\
0=&\omega h_{02}'+k f h_{12}'.
\end{split}
\end{equation}
Where primes denote derivatives with respect to the radial direction $z$.

An important ingredient in our derivation is parity symmetry. The equations are invariant under the transformation
\begin{equation}\label{paritytrans}
k\longrightarrow -k, \ \ h_{12}\longrightarrow - h_{12}.
\end{equation}
Therefore, for every solution $h_{02}$, $h_{12}$ of the equations with frequency $\omega$ and momentum $k$, there is another solution for the opposite momentum with the same radial profile, up to the overall sign in $h_{12}$
\begin{equation}
 \tilde{h}_{02}(\omega,-k,z)=h_{02}(\omega,k,z),\ \ \tilde{h}_{12}(\omega,-k,z)=-h_{12}(\omega,k,z).
\end{equation}
Note that generically introducing sources in the field theory will break parity, this will be reflected in the boundary conditions $h_{12}(\omega,k,z=0)\neq -h_{12}(\omega,-k,z=0)$ and $h_{02}(\omega,k,z=0)\neq h_{02}(\omega,-k,z=0)$. Nonetheless, the spectrum is determined by normalizable solutions $h_{12}(z=0)=h_{02}(z=0)=0$, which will show parity invariance.

\subsection{Probability current and parity}

The equations of motion \eqref{eoms} can be cast in the form of coupled Schr\"oedinger equations (plus a constraint) by changing to the following variables 
\begin{equation}
h_{02}= z \psi_{0},\ \ h_{12}=\frac{z}{\sqrt{f}} \psi_{1}.
\end{equation}
Then, the dynamical equations have the form
\begin{equation}
\begin{split}
0=& \psi_{1}''-\cV_{1}\psi_{1}+\frac{\omega k}{f^{3/2}}\psi_{0},\\
0=& \psi_{0}''-\cV_{0} \psi_{0}-\frac{\omega k}{f^{3/2}}\psi_{1}.\\
\end{split}
\end{equation}
Where
\begin{equation}
\cV_{1}=-\frac{\omega^2}{f^2}+\frac{f'}{f z}-\frac{\left(f'\right)^2}{4
	f^2}+\frac{3}{f z^2}-\frac{1}{z^2},\ \ \cV_{0}=\frac{k^2}{f}+\frac{2}{z^2}.
\end{equation}

For $k=0$ the two modes decouple and one can naturally define probability currents for each of the fluctuations (bar denotes complex conjugation)
\begin{equation}
j_{0,1}=\overline{\psi}_{0,1} {\psi_{0,1}}'-{\overline{\psi}_{0,1}}'\psi_{0,1}.
\end{equation}
Since the potentials are real, these currents  are independent of the radial coordinate
\begin{equation}
\frac{d}{dz} j_{0,1}=\overline{\psi}_{0,1} {\psi_{0,1}}''-{\overline{\psi}_{0,1}}''\psi_{0,1}=\overline{\psi}_{0,1} (\cV_{0,1}  \psi_{0,1})-(\cV_{0,1}  \overline{\psi}_{0,1})\psi_{0,1}=0.
\end{equation}
However, when the momentum is non-zero the currents $j_0$ and $j_1$ are not independent of the radial direction anymore. Instead, we find
\begin{equation}
\frac{d}{dz} j_1=-\frac{\omega k}{f^{3/2}}(\overline{\psi}_1\psi_0-\overline{\psi}_0\psi_1),\ \ \frac{d}{dz} j_0=\frac{\omega k}{f^{3/2}}(\overline{\psi}_0\psi_1-\overline{\psi}_1\psi_0).
\end{equation}
Even though separately the radial derivative of each current is non-zero we see that they are proportional to the same function. The combination $J=j_1-j_0$ is actually independent of the radial direction. Note that $J$ is invariant under the parity transformation \eqref{paritytrans}, since it depends quadratically on $h_{12}$ and the radial profile of the fluctuations does not change. We will use this fact to derive the generalized Ward identity by comparing the value of the current $J$ at the horizon and at the boundary. 

To leading order, the ingoing solutions near the horizon $z\to 1$ can be expanded as
\begin{equation}
\begin{split}
\psi_1= &(1-z)^{\frac{1}{2}-i\frac{\omega}{3}} \left( \cA_1+\cdots\right)+\sqrt{1-z}\left(-\frac{\sqrt{3}k}{\omega}\cB+\cdots \right),\\ 
\psi_0= & (1-z)^{1-i\frac{\omega}{3}} \left( \frac{i\sqrt{3}k}{3-i\omega}\cA_1+\cdots\right)+\cB\left(1+(1-z)+\cdots\right). 
\end{split}
\end{equation}
The solution for $\psi_0$ when the momentum is zero and the two modes are decoupled is the one with coefficient $\cB$. In that situation, there are no ingoing solutions for the vector mode. This solution actually does not contribute to the current, whose value at the horizon is
\begin{equation}\label{eq:JH}
J_H=i\frac{2}{3} \omega \left|\cA_1\right|^2.
\end{equation} 
Invariance of the current under parity implies that $J_H(-\cA_1,-k)=J_H(\cA_1,k)$. But
\begin{equation}\label{eq:JH2}
J_H(-\cA_1,-k)=i\frac{2}{3} \omega \left|-\cA_1(-k)\right|^2=i\frac{2}{3} \omega \left|\cA_1(-k)\right|^2=J_H(\cA_1,-k).
\end{equation} 
Therefore, $J_H(k)=J_H(-k)$ and since $J$ is independent of the radial coordinate, we deduce that
\begin{equation}\label{Jcond}
J(k)=J(-k).
\end{equation}

\subsection{Ward identities}

To leading order, the solutions near the boundary $z\to 0$ can be expanded as
\begin{equation} \label{psiboun}
\psi_{0,1}=\frac{1}{z}\left(H_{0,1}^{(0)}+H_{0,1}^{(2)} z^2+\cdots\right)+z^2\left(T_{0,1}^{(0)}+\cdots\right).
\end{equation}
The coefficients of the non-normalizable modes $H_{0,1}^{(0)}$ are proportional to the Fourier transform of the sources of the energy-momentum tensor, i.e. the boundary metric 
\begin{equation}
g_{\mu\nu}^{(0)}=\eta_{\mu\nu}+\delta g_{\mu\nu}^{(0)}.
\end{equation}
The coefficients of the normalizable modes $T_{0,1}^{(0)}$ are related to the Fourier transform of the expectation value in the dual field theory 
\begin{equation}
\vev{T_{\mu\nu}}=\vev{T_{\mu\nu}}_{\text{thermal}}+\delta \vev{T_{\mu\nu}}.
\end{equation}
The exact relation follows from holographic renormalization \cite{deHaro:2000xn}
\begin{equation}\label{bounddict}
\delta g_{a2}^{(0)}=-H_{a}^{(0)}, \, a=0,1. \ \ \delta \vev{T_{02}}=- \frac{3L^2}{16\pi G_N}T_0^{(0)},\ \ \ \ \delta\vev{T_{12}}=- \frac{3L^2}{16\pi G_N}\left[T_1^{(0)}+\frac{1}{2}H_1^{(0)}\right].
\end{equation}
The two-point correlation functions of the energy-momentum tensor can be computed by doing a variation of the expectation value with respect to the source
\begin{equation}\label{corr}
\Gamma_{a2b2}=\frac{\partial  \vev{T_{a2}}}{\partial g^{b2}}\Big|_{\delta g_{\mu\nu}=0}.
\end{equation}
Note that the normalizable solution is not independent, ingoing boundary conditions at the horizon impose a relation between the boundary normalizable and non-normalizable solutions, with some coefficients depending on the frequency and the momentum
\begin{equation}
\begin{split}
T_1^{(0)}=& C_{11}(\omega,k) H_1^{(0)}+C_{10}(\omega,k) H_0^{(0)},\\
T_0^{(0)}=& C_{01}(\omega,k) H_1^{(0)}+C_{00}(\omega,k) H_0^{(0)}.
\end{split}
\end{equation}
From \eqref{bounddict} and \eqref{corr}, the correlation functions associated to the shear and transverse thermal conductivity are
\begin{equation}
\begin{split}
&\Gamma_{0202}=- \frac{3L^2}{16\pi G_N} C_{00}, \ \ \Gamma_{1202}=- \frac{3L^2}{16\pi G_N}C_{10},\\
&\Gamma_{0212}=- \frac{3L^2}{16\pi G_N} C_{01}, \ \ \Gamma_{1212}=- \frac{3L^2}{16\pi G_N}\left[C_{11}+\frac{1}{2}\right].
\end{split}
\end{equation}

The coefficients $H_{0,1}^{(2)}$ are fixed by the corresponding dynamical equations in \eqref{eoms}. If we also take into account the constraint equation, we find the following conditions
\begin{equation}\label{eq:boundcoeff}
\begin{split}
H_1^{(2)} =& \frac{\omega^2}{2}H_1^{(0)}+\frac{\omega k}{2}H_0^{(0)},\\
T_0^{(0)}=& -\frac{k}{2\omega} H_1^{(0)}-\frac{k}{\omega} T_1^{(0)},\\
H_0^{(2)}=& -\frac{k}{\omega} H_1^{(2)}= -\frac{\omega k}{2}H_1^{(0)}-\frac{ k^2}{2}H_0^{(0)}.
\end{split}
\end{equation}
Using all these results, the probability current evaluated at the boundary has the following form 
\begin{equation}\label{eq:JB}
J_B=3\left({\overline{H}_1^{(0)}} T_1^{(0)}-H_1^{(0)} {\overline{T}_1^{(0)}} \right)+\frac{3k}{2\omega}\left({\overline{H}_0^{(0)}} H_1^{(0)}-H_0^{(0)} {\overline{H}_1^{(0)}} \right)+\frac{3k}{\omega}\left({\overline{H}_0^{(0)}} T_1^{(0)}-H_0^{(0)} {\overline{T}_1^{(0)}} \right).
\end{equation} 
We see that indeed $J_B$ is invariant under parity, since $J_B(-H_1,-T_1,-k)=J_B(H_1,T_1,k)$. However, in order for the condition \eqref{Jcond} to hold, an additional constraint should be satisfied (``even'' and ``odd'' is respect to $k\to -k$) 
\begin{equation}\label{constJ}
\omega \left({\overline{H}_1^{(0)}} T_1^{(0)}-H_1^{(0)} {\overline{T}_1^{(0)}} \right)_{\text{odd}}+\frac{k}{2}\left({\overline{H}_0^{(0)}} H_1^{(0)}-H_0^{(0)} {\overline{H}_1^{(0)}} \right)+k\left({\overline{H}_0^{(0)}} T_1^{(0)}-H_0^{(0)} {\overline{T}_1^{(0)}} \right)_{\text{even}}=0.
\end{equation}
We will use this relation to derive Ward identities for correlators in the dual field theory.

Using \eqref{eq:boundcoeff}, we see that not all those coefficients are independent, but
\begin{equation}\label{eq:gvv}
\omega C_{01}(\omega,k)+k C_{11}(\omega,k)+\frac{k}{2}=0,\ \ \omega C_{00}(\omega,k)+kC_{10}(\omega,k)=0.
\end{equation}
This implies the following Ward identities
\begin{equation}\label{ward1}
\omega \Gamma_{0212}+k \Gamma_{1212}=0, \ \ \omega \Gamma_{0202}+k \Gamma_{1202}=0,
\end{equation}
which correspond to the usual conservation of the energy-momentum tensor.

On the other hand, \eqref{constJ} for arbitrary sources leads to the conditions
\begin{align}
\left(C_{11}-\overline{C}_{11}\right)_{\text{odd}} &=0,\\
\omega C_{10\,\text{odd}}-\frac{k}{2}-k\, \overline{C}_{11,\text{even}} &=0.
\end{align}
This implies the following Ward identities
\begin{equation}\label{ward2}
\left(\Gamma_{1212}-\overline{\Gamma}_{1212}\right)_{\text{odd}}=0, \ \ \omega \Gamma_{1202\,\text{odd}}-k \overline{\Gamma}_{1212\,\text{even}}=0.
\end{equation}
Combining the two second identities in \eqref{ward1} and \eqref{ward2}, we get
\begin{equation}\label{wardshear}
\left[\omega^2 \Gamma_{0202}+k^2 \overline{\Gamma}_{1212}\right]_{\text{even}}=0.
\end{equation}
From this expression we can derive the relation between shear viscosity and transverse thermal conductivity.

\section{Bulk identity in a non-CFT}\label{bulkid}

We would like now to study a more involved case, the relation between bulk viscosity and longitudinal thermal conductivity. Although the derivation is similar to the one we have used for the shear identity in a CFT, there are some features that cannot be easily generalized, in particular finding a constant probability current for a larger number of coupled fluctuations. We will show how this can be done using a particular example.

In a CFT the bulk viscosity is zero, so we should introduce a breaking of conformal invariance. This can be achieved by introducing additional couplings for relevant operators or giving them an expectation value. On the gravity side, this is translated into turning on scalar fields. In the simplest scenario there will be just one scalar coupled to Einstein gravity  
\begin{equation}\label{acmat}
S=\frac{1}{16\pi G_N}\int d^4 x\, \sqrt{-g}\left(R-(\partial\phi)^2-2 V(\phi)\right).
\end{equation}
The potential has a critical point at $\phi=0$ corresponding to $AdS$ with $\Lambda= V(0)$. In order for the dual operator $\cO$ to be relevant, the mass of the scalar field should be negative $m^2L^2=\partial^2 V(0)<0$, so the critical point is a maximum of the potential. The mass of the field is related to the conformal dimension of the dual operator $\Delta$ as $m^2L^2=\Delta(\Delta-3)$.

The equations of motion are
\bea \label{geoms2}
R_{MN}-\frac{1}{2}R g_{MN}&=&  \partial_M \phi \partial_N \phi -\frac{1}{2}g_{MN}\left(\pd_K \phi \partial^K \phi +2V(\phi) \right) \,,\\
\label{keoms2}
0&=& \Box\phi -\partial V(\phi)\,.
\nonumber
\eea
We will consider a generic background black brane solution with the scalar field turned on 
\begin{equation} \label{gbulk}
ds^2=dr^2+e^{2A(r)}\left(-e^{2B(r)} dx_0^2+dx_1^2+dx_2^2\right),\ \ \phi=\phi_0(r).
\end{equation}
The boundary is at $r\to \infty$,  where the solution is asymptotically $AdS$. Asymptotically close to the boundary the scalar field and the blackening function of the metric vanish $\phi_0, B\to 0$, while the warp factor becomes linear in the radial coordinate $A(r)\simeq \frac{r}{L}$. The horizon is at $r=r_H$, where the $g_{00}$ component of the metric vanishes $B(r) \simeq \log(r-r_H)$.  For convenience we will set $L=1$ in the calculations, so all dimensionful quantities are given in units of the $AdS$ radius. We will restore the dependence on $L$ in the final expressions for one-point functions and correlators. To leading order,\footnote{We are presenting the expansions as if $3/2<\Delta <5/2$ , but they are valid for any $1/2<\Delta <3$, except for special values, when $2\Delta -3$ is an integer. } the expansion of the solutions close to the boundary are
\begin{equation}
\begin{split}
&A\sim r -\frac{\lambda^2}{8}e^{-2(3-\Delta)r}-\frac{1}{9}\left( 3B_0-\Delta(\Delta-3)\lambda v \right)e^{-3 r}+\cdots,\\ 
&B\sim e^{-3 r} B_0+\cdots ,\ \  \phi_0 \sim \lambda e^{-(3-\Delta) r}+v e^{-\Delta r}+\cdots.
\end{split}
\end{equation}
The coefficients $\lambda$ and $v$ are proportional to the source and the expectation value of the dual scalar operator respectively. The coefficient $B_0$ is proportional to the thermal contribution to the energy density. We have computed the renormalized expectation values in Appendix~\ref{apren} using the holographic renormalization procedure.
 We find that the total energy density $\ep$ and pressure $P$ are 
\begin{equation} \label{energy}
	\begin{split}
		\ep = \vev{T_{00}}=& \frac{1}{8\pi G_N L}\left[-2B_0 -\frac{1}{3} (\Delta -3)(3-2\Delta) \lambda  v\right],\\
		P = \vev{T_{ii}}=& \frac{1}{8\pi G_N L}\left[-B_0+\frac{1}{3}(\Delta-3)(3-2\Delta)\lambda v\right],
	\end{split}
\end{equation}
while the expectation value of the dual scalar operator is
\begin{equation}\label{vevO}
\vev{\cO}= \frac{\mu^{\Delta-3}}{8\pi G_N L} (3-2\Delta) v.
\end{equation}
Where $\mu$ is an arbitrary scale that enters in the definition of the source for the dual operator $\lambda =\mu^{\Delta-3}J^{(0)}$.

The trace of the energy-momentum tensor satisfies the Ward identity
\begin{equation}
	\vev{T^\mu_\mu}=\frac{1}{8\pi G_NL}(\Delta -3) (3-2 \Delta) \lambda  v =(\Delta-3)J^{(0)} \vev{\cO}.
\end{equation}

In order to compute correlation functions in the dual field theory we follow the usual analysis of linearized fluctuations of the metric and scalar field $\delta g_{MN}$, $\delta\phi$. We will work in the radial gauge $\delta g_{rM}=0$ and expand in plane waves along the field theory directions
\begin{equation} \label{fluctuation}
\delta g_{\mu\nu}=\int \frac{d^3 k}{(2\pi)^3 } e^{ik_\mu x^\mu} h_{\mu\nu}(r),\ \ \delta\phi=e\int \frac{d^3 k}{(2\pi)^3 } e^{ik_\mu x^\mu}\varphi(r).
\end{equation}
It is possible to repeat the derivation of the shear identity in this background by turning on only the $h_{12}$ and $h_{02}$ components of the metric. The structure of the equations is the same, with the only difference being that when the equations are written in the Schr\"oedinger form the potentials depend on the background scalar.

For the present analysis we will turn on the minimal set of modes of the metric coupled to the scalar $\varphi$: $h_{00}$, $h_{01}$, $h_{11}$ and $h_{22}$ and we will fix the momentum to be $k^\mu =(\omega,k,0)$ without loss of generality. It will be convenient to use a different basis of modes
\begin{equation} \label{changes}
\begin{split}
&y_1=\frac{1}{2}e^{-\frac{A}{2}}\left(e^{-2B}h_{00}+h_{11}+h_{22}\right),\\
&y_2=\frac{1}{2}e^{A}\left(-e^{-2 B} h_{00}+h_{11}+h_{22}\right),\\
&y_3 = e^{\frac{3}{2}A}\varphi,\\
&y_4=\frac{1}{2}e^{-\frac{A}{2}}\left(h_{11}-h_{22}\right),\\
&y_5=e^{-\frac{A}{2}-2B} h_{01}.
\end{split}
\end{equation}
The dynamical equations of motion and the constraints are
\begin{equation}\label{eqsy}
y_i''+a_{ij}\, y_j'+b_{ij}\, y_j=0, \ \ c_i^a\,y_i'+d_i^a\, y_i=0, \ a=1,2,3.
\end{equation}
Where the coefficients are given in the Appendix~\ref{app:fluceqs}. The structure is such that, for $i\neq 5$
\begin{equation}
a_{5i}=a_{i5}=0, \ \ b_{i5}, b_{5i}\propto k, \ \ c^1_5, c^3_5, c^2_i\propto k,\ \ d^5_5, d^3_5, d^2_i \propto k.
\end{equation}
While all the other coefficients are proportional to even powers of $k$. One can easily check that the equations are invariant under the parity transformation
\begin{equation}
k\longrightarrow -k, \ \ h_{01}\longrightarrow -h_{01} \ \ (y_5 \longrightarrow- y_5). 
\end{equation}

\subsection{Constructing a probability current}\label{probcurr}

For the shear modes it was quite easy to construct a constant probability current. This was actually possible because the structure of the equations is quite special. The existence of a constant probability current was pointed out before in other simple cases, such as a probe scalar field in the BTZ black hole  \cite{Son:2002sd} and for longitudinal fluctuations of a gauge field in a charged asymptotically $AdS_4$ black hole \cite{Amado:2009ts}. In those cases it is roughly proportional to the on-shell action. As we will see, in the more complicated case of a scalar field coupled to the metric, the special structure of the vector modes is absent and in order to construct a constant current we need to introduce additional ingredients. Let us mention that it is possible to define in general a conserved `symplectic current' $w^\mu$ \cite{Lee:1990:LSC,Wald:1999wa} that is useful to prove the conservation of Noether charges in the bulk and the first law of thermodynamics in holography (see \cite{Papadimitriou:2005ii,Papadimitriou:2010as}). Similarly to the probability current, it is a bilinear functional of the fluctuations of the fields $w^\mu=w^\mu(\delta_1 \Phi,\delta_2\Phi)$. However, in contrast to the probability current defined from the on-shell action, it vanishes for $\delta_1\Phi=\delta_2\Phi$. It would be interesting to see if the probability current we define below and the symplectic current are related.

Let us first consider zero momentum $k=0$ with the background scalar field turned off $\phi_0=0$. The dynamical equations take a simpler form, with all the modes decoupled except for $y_1$ and $y_2$. For each of the decoupled modes the equations are
\begin{equation}
y_i''+a_{ii} y_i'+b_{ii}y_i=0,\ i=3,4,5.
\end{equation}
We can define new variables such that the equations take the form of Schr\"oedinger equations. The new variables are
\begin{equation}
y_i(r)=e^{-\frac{1}{2}\int^r a_{ii}}\psi_i(r),
\end{equation}
and the equations become
\begin{equation}
\psi_i''-V_i\psi_i=0,\ \ V_i=-b_{ii}+\frac{1}{2}a_{ii}'-\frac{1}{4}a_{ii}^2.
\end{equation}
Therefore, there is a constant probability current for each of these modes
\begin{equation}
j_i={\overline{\psi}_i}'\psi_i-\overline{\psi}_i{\psi_i}'.
\end{equation}

For the coupled modes, the coefficients of the dynamical equations are 
\begin{equation}
\arraycolsep=6pt
a_{lm}=\left(
\begin{array}{cc}
B' & e^{-\frac{3 A}{2}} \left(A'-B'\right)  \\
0 & 2 B' 
\end{array}
\right), 
\end{equation}
and
\begin{equation}
b_{lm}=\left(
\begin{array}{ccccc}
-\frac{3}{4} A' \left(3 A'+4 B'\right) & -3 e^{-\frac{3 A}{2}} A'\left(A'-B'\right)  \\
e^{-\frac{A}{2}-2 B} \omega^2 &  e^{-2 A-2B}  \omega^2-9 A'\left(A'+B'\right)   
\end{array}
\right)\,, \quad l,m = 1,2\,.\\
\end{equation}
We can also put this equation in Schr\"oedinger form by defining new variables
\begin{equation}
y_l=\Omega_{lm}\psi_m, \ \ l,m=1,2.
\end{equation}
The matrix $\Omega$ has to satisfy the differential equation
\begin{equation}
\Omega ' =-\frac{a}{2}\Omega,
\end{equation}
for which there is a formal solution
\begin{equation}
\Omega=\mathbb{1}+\sum_{n=1}^\infty \frac{(-1)^n}{2^n} \int^r dr_1 \int^{r_1} dr_2 \cdots \int^{r_{n-1}} dr_n\,  a(r_1) a(r_2) \cdots  a(r_n).
\end{equation}
The equations become, in matricial form
\begin{equation}\label{schreq2}
\psi''-\cV \psi=0,\ \ \cV=\Omega^{-1} \left[\frac{a^2}{4}+\frac{a'}{2}-b \right]\Omega.
\end{equation}
However, in this case we cannot construct a simple probability current. Na\"{\i}vely the general form would be
\begin{equation}
J={\psi^\dagger}'  Q\psi -\psi^\dagger Q^\dagger {\psi}',
\end{equation}
with $Q$ a constant matrix. In order for the current to be constant it is necessary that the matrix $Q$ satisfies the algebraic equations
\begin{equation}
\cV^\dagger Q- Q^\dagger\cV=0, \ \ Q= Q^\dagger.
\end{equation}
For the shear mode this was the case because $\cV$ can be expanded in Pauli matrices $\{ \mathbb{1},\sigma^3, i\sigma^2\}$ with real coefficents, so the algebraic relations are satisfied for $Q= \sigma^3$.  However, for the scalar modes there is also a term in the potential proportional to $\sigma^1$, so the algebraic constraints cannot be satisfied in general.

A possible way to generalize the probability current would be to allow non-constant coefficients, and define the current as
\begin{equation}
J= {\psi^\dagger}' \cA {\psi}' +{\psi^\dagger}' \cB {\psi}-{\psi^\dagger} \cB^\dagger {\psi}'+{\psi^\dagger} \cC {\psi}, 
\end{equation}
with $\cA^\dagger=-\cA$, $\cC^\dagger=-\cC$. The condition that the current is constant $J'=0$ together with the equations of motion \eqref{schreq2} gives differential equations for the coefficents.
A current of this form could also be defined for the original system of equations (even if $k\neq 0$ and $\phi_0\neq 0$), without having to write the equations in Schr\"oedinger form. The current would be
\begin{equation}\label{generalJ}
J= {y^\dagger}' \cA {y}' +{y^\dagger}' \cB {y}-{y^\dagger} \cB^\dagger {y}'+{y^\dagger} \cC {y}. 
\end{equation}
Rather than constructing $J$ in this way and solving the differential equations for the coefficients, we will try a different approach. We will add additional fields so the probability current becomes a Noether current with known coefficients and then we will fix the boundary conditions of the auxiliary fields in terms of the boundary conditions of the original modes.

First let us introduce a matrix $K$ such that the equations can be written in the form 
\begin{equation}
K^{-1}\left( K y'\right)'+b y=0,\ \ K^{-1}K'=a\ \ \Rightarrow \ \ K'=Ka.
\end{equation}
A formal solution is
\begin{equation}
K=\mathbb{1}+\sum_{n=1}^\infty \int^r dr_1 \int^{r_1} dr_2 \cdots \int^{r_{n-1}} dr_n\, a(r_n)\cdots a(r_2) \, a(r_1).
\end{equation}
If we multiply by $K$ on the left we get
\begin{equation}\label{eqsy2}
\left( K y'\right)'+K b y=0.
\end{equation}
We can derive this from a Lagrangian by introducing new fields $\eta$. The number of auxiliary fields is the same as the number of original fluctuations and can be grouped in a vector of the same length. The Lagrangian that gives the equations for $y$ is
\begin{equation}
\cL=(\eta^\dagger)' K y'-\eta^\dagger K b y+(y^\dagger)' K^\dagger \eta'-y^\dagger b^\dagger K^\dagger \eta. 
\end{equation}
\begin{equation}\label{eqeta}
\left( K^\dagger \eta'\right)'+b^\dagger K^\dagger \eta=0 \,.
\end{equation}
Note also that the equations are invariant under the parity transformation
\begin{equation}
k\longrightarrow -k, \ \ \eta_5\longrightarrow -\eta_5,
\end{equation}
 this will be important in the derivation of the new identities.

The equations for $y$ and $\eta$ become the same if $K=K^\dagger$ and $K b=b^\dagger K$, in which case one can set $\eta=y$ and $\cL$  can be used as a Lagrangian for the original system of equations.

The action of the extended system has a $U(1)$ global symmetry
\begin{equation}
y\longrightarrow e^{i\alpha}y, \ \ \eta\longrightarrow e^{i\alpha}\eta,
\end{equation}
whose (anti-Hermitian) Noether current is
\begin{equation}\label{noetherJ}
J=(\eta^\dagger)' Ky -\eta^\dagger K y'+(y^\dagger)' K^\dagger \eta-y^\dagger K^\dagger \eta'.
\end{equation}
The equations of motion imply that $J'=0$. The current is invariant under the full parity symmetry acting on both $y$ and $\eta$.

\subsubsection{Current at the boundary}

Our first goal is to compute the probability current at the boundary. For simplicity we will restrict to a quadratic potential for the scalar field $V(\phi)=\frac{1}{2}m^2\phi^2$, with $m^2<0$ but above the Breitenlohner-Freedman bound, as is appropriate for a field dual to a relevant operator.  For arbitrary potentials $V(\phi)$ with a maximum at $\phi=0$ we have checked that there are no qualitative changes in the boundary expansions, although coefficients do depend on third and fourth derivatives of the potential. 

The expansions of the background and the matrix $K$ can be found in Appendix~\ref{app:backexpb}, and the one for the auxiliary fields in Appendix~\ref{app:flucexpb}. Since the equations are second order, there are in principle two independent solutions for each of the $y_i$ and $\eta_i$. One corresponds to the non-normalizable solution, which for the original fluctuations maps to the metric or to a source for the scalar operator in the dual field theory. The other solution is normalizable and for the original fluctuations maps to the expectation value of the energy-momentum tensor and the scalar operator. Let us compare the leading terms of each of the independent solutions in the expansions of the auxiliary fields to those of the original fluctuations.
\allowdisplaybreaks
\begin{flalign*}
&y_1 \sim e^{\frac{3}{2} r} y_1^{(0)}+e^{-\frac{3}{2} r} y_1^{(3)}, \\
&\eta_1 \sim  e^{\frac{3}{2} r}\eta_1^{(0)} + e^{-\frac{3}{2} r}\eta_1^{(3)} -\frac{1}{8}(k^2+2\omega^2)e^{\frac{5}{2} r} \eta_2^{(0)}+\cdots,\\ &\\
&y_2 \sim e^{3 r} y_2^{(0)} + e^{-3 r}y_2^{(6)}, \\ 
&\eta_2 \sim e^{3 r} \eta_2^{(0)}+e^{-3 r} \eta_2^{(6)},\\ & \\
&y_3 \sim e^{-\left(\frac{3}{2}-\Delta\right) r} y_3^{(3-\Delta)}+e^{\left(\frac{3}{2} -\Delta\right) r} y_3^{(\Delta)}, \\ 
&\eta_3 \sim e^{-\left(\frac{3}{2}-\Delta\right) r} \eta_3^{(3-\Delta)}+e^{\left(\frac{3}{2} -\Delta\right) r} \eta_3^{(\Delta)}+\frac{\Delta-3}{2} e^{\left(\frac{3}{2}+\Delta \right) r}\lambda \eta_2^{(0)}+\cdots,\\ & \\
&y_4 \sim e^{\frac{3}{2} r} y_4^{(0)}+e^{-\frac{3}{2} r} y_4^{(3)}, \\ 
&\eta_4 \sim e^{\frac{3}{2} r} \eta_4^{(0)}+e^{-\frac{3}{2} r} \eta_4^{(3)}-\frac{k^2}{4}e^{\frac{5}{2}r} \eta_2^{(0)}+\cdots,\\ & \\
&y_5 \sim e^{\frac{3}{2} r} y_5^{(0)}+e^{-\frac{3}{2} r} y_5^{(3)}, \\ 
&\eta_5 \sim e^{\frac{3}{2} r} \eta_5^{(0)}+e^{-\frac{3}{2} r} \eta_5^{(3)}-\frac{k\omega}{2}e^{\frac{5}{2}r} \eta_2^{(0)}+\cdots.
\end{flalign*} 
The independent terms are the same, but the leading terms in the auxiliary fields start with a larger exponent due to the mixing with $\eta_2$. This can be understood as follows, close to the boundary the coefficients $a\to \mathbb{0}$, making $K\to \mathbb{1}$ and $b$ becomes diagonal. If all the modes had the same asymptotics, then we will be in the case where we can set $\eta=y$. However, this is not exactly true because $\eta_2$ grows faster than the other modes and even though the off-diagonal components of $b$ and $a$ go to zero at the boundary, they do not decay fast enough to avoid  the mixing. Nonetheless, while we cannot impose the condition $\eta=y$, we can fix some relation between the leading coefficients of the independent solutions. There is an ambiguity in this choice, since different combinations may be formed. The simplest option is simply to match the leading coefficients of each of the independent solutions for $y$ with the leading coefficients of the independent solutions for $\eta$ 
\begin{equation}\label{yetamap}
\arraycolsep=1.4pt\def\arraystretch{2.2}
\begin{array}{ll}
\eta_i^{(0)}=y_i^{(0)}, \ \ &\eta_i^{(3)}=y_i^{(3)},\ \   i=1,4,5,\\
\eta_2^{(0)}=y_2^{(0)}, \ \ &\eta_2^{(6)}=-y_2^{(6)},\\
\eta_3^{(3-\Delta)}=y_3^{(3-\Delta)},\ \ &\eta_3^{(\Delta)}=y_3^{(\Delta)}.
\end{array}
\end{equation}

This fixes completely the auxiliary modes in terms of the original fluctuations.\footnote{The choice of sign for $\eta_2^{(6)}$ gives simpler expressions.} We can group the non-normalizable coefficients in a vector $H$ and the normalizable coefficients in another vector $T$,
\begin{equation}
H^T =\left(\begin{array}{ccccc} y_1^{(0)} & y_2^{(0)} & y_3^{(3-\Delta)} & y_4^{(0)} & y_5^{(0)} \end{array} \right), \ T^T=\left(\begin{array}{ccccc} y_1^{(3)} & y_2^{(6)} & y_3^{(\Delta)} & y_4^{(3)} & y_5^{(3)} \end{array} \right).
\end{equation}
In the following we will use the vector components $H_i$ and $T_i$ to refer to the normalizable and non-normalizable coefficients. The current evaluated at the boundary has the form
\begin{equation}
J_B= H^\dagger \cC H + H^\dagger \cD T -T^\dagger \cD^\dagger H,
\end{equation}
where $\cC^\dagger = -\cC$. The non-zero elements of $\cC$ are
\begin{equation}
\cC_{12}=-\cC_{21},\ \cC_{13}=-\cC_{31}, \ \cC_{23}=-\cC_{32}, \ \cC_{24}=-\cC_{42}, \ \cC_{25}=-\cC_{52}.
\end{equation}
They have even powers of $k$ except for $\cC_{25}$, which is proportional to odd powers. The explicit value of the coefficients is given in  Appendix~\ref{app:currcoefs}.
The non-zero elements of $\cD$ are
\begin{equation}
\cD_{11}, \ \cD_{22},\ \cD_{33},\ \cD_{44}, \ \cD_{55},\ \cD_{13},\ \cD_{21},\ \cD_{23},
\end{equation}
all of which only have even powers of $k$. The explicit value is also in Appendix~\ref{app:currcoefs}. One can easily check that current is explicitly invariant under the parity transformation
\begin{equation}
k\longrightarrow -k, \ \ H_5\longrightarrow -H_5,\ \ T_5\longrightarrow -T_5.
\end{equation}
In order to compute correlators we impose ingoing boundary conditions at the horizon. At the boundary this means that the coefficients of the normalizable solutions are not independent, but they are proportional to the coefficients of the non-normalizable solutions:
\begin{equation}\label{TG}
T_i = C_{ij}(\omega,k) H_j,
\end{equation}
or, in matrix notation, $T=CH$. Then, the current evaluated at the boundary can be written as
\begin{equation}\label{JB}
J_B= H^\dagger \cG H,\ \ \cG=\cC+\cD C-C^\dagger\cD^\dagger.
\end{equation}

\subsubsection{Current at the horizon}
\label{sec:hocu}

We now proceed to compute the probability current at the horizon. The background is taken to be regular at the horizon, with $g_{00}$ having a simple zero at $r=r_H$. 
\begin{equation} \label{backhor}
A= A_H +O((r-r_H)^2)\,, B= \log \left(r-r_H \right)+ B_H +O((r-r_H)^2)\,, \ \phi= \phi_H +O((r-r_H)^2).
\end{equation}
To leading order, the matrix $K$ close to the horizon takes the following form
\begin{equation}
\arraycolsep=6pt
K=(r-r_H)\left(
\begin{array}{lllll}
K_{11}^H   & K_{12}^H(r-r_H)+ e^{-\frac{3}{2}A_H} K_{11}^H  & & & \\
& K_{22}^H( r-r_H) & & & \\
& K_{32}^H (r-r_H) & K^H_{33} & & \\
& & & K_{44}^H  & \\
& & & & K_{55}^H (r-r_H)^2
\end{array}
\right).
\end{equation}
The expansions of the background and the matrix $K$ can be found in Appendices~\ref{app:backexph}. 
The coefficients $K_{ij}^H$ are not determined by the expansion close to the horizon. Their value can be determined by solving the differential equation for $K$ imposing the condition at the boundary that $K\to \mathbb{1}$. 

Since the equations for the fluctuations are of second order, there are in principle two independent solutions for each of the $y_i$ and $\eta_i$, generically of the form $\sim (r-r_H)^\alpha$. The exponent $\alpha$ can be complex. For the fluctuations $y_i$ we impose regularity (if $\alpha$ is real) or ingoing boundary conditions (if $\alpha$ is complex). Since we have already imposed the conditions \eqref{yetamap}, there is no freedom left to fix the behavior of the auxiliary fields $\eta_i$ at the horizon. 
The leading order terms of each of the independent solutions are
\allowdisplaybreaks \label{fluchor}
\begin{align*} 
&y_1 \sim y_1^H , \\
&\eta_1\sim  (r-r_H)^{-i c_H\, \omega} \eta_1^H +(r-r_H)^{i c_H\, \omega} \tilde{\eta}_1^H,  \\ & \\
&y_2 \sim (r-r_H)^{-ic_H\, \omega} y_2^H, \\ 
&\eta_2 \sim \eta_2^H +\frac{\tilde{\eta}_2^H}{r-r_H},\\ & \\
&y_3 \sim (r-r_H)^{-ic_H\, \omega} y_3^H,\\ 
&\eta_3 \sim  (r-r_H)^{-i c_H\, \omega} \eta_3^H +(r-r_H)^{i c_H\, \omega} \tilde{\eta}_3^H,\\ & \\
&y_4 \sim (r-r_H)^{-i c_H\,\omega} y_4^H, \\ 
&\eta_4 \sim (r-r_H)^{-i c_H\, \omega} \eta_4^H +(r-r_H)^{i c_H\, \omega} \tilde{\eta}_4^H,\\ & \\
&y_5 \sim y_5^H  \\ 
&\eta_5 \sim \eta_5^H+\frac{\tilde{\eta}_5^H}{(r-r_H)^2}.
\end{align*} 
Where we have defined $c_H=e^{-(A_H+B_H)}$. All the fluctuations are actually mixed, the expansion of the fluctuations and the auxiliary fields can be found in Appendix~\ref{app:flucexph}.

Let us group the coefficients of the solutions in the vectors $y_H$, $\eta_H$ and $\tilde{\eta}_H$ with components
\begin{equation}
(y_H)_i =y_i^H, \ \ (\eta_H)_i=\eta_i ^H,\ \ (\tilde{\eta}_H)_i=\tilde{\eta}_i^H.
\end{equation}
The probability current evaluated at the horizon takes the form
\begin{equation}
J_H= \eta_H^\dagger \cM y_H-y_H^\dagger \cM^\dagger \eta_H+\tilde{\eta}_H^\dagger \cN y_H-y_H^\dagger \cN^\dagger \tilde{\eta}_H.
\end{equation}
Where the non-zero entries of each matrix are
\begin{equation}
\begin{split}
&\cM_{12} =2e^{-\frac{3}{2}A_H} (ic_H\, \omega-1) K_{11}^H, \ \ \cM_{33}=2i c_H\, \omega K_{33}^H,\ \ \cM_{44}=2i c_H\, \omega K_{44}^H,\\
&\cN_{21}=e^{\frac{3}{2}A_H} K_{22}^H, \ \ \cN_{55}=-2 K_{55}^H.
\end{split}
\end{equation}

When we solve the linear equations of motion, we can write a general solution in terms of the boundary values using a boundary-to-bulk propagator
\begin{equation}
y_i= G_{ij}(r,\omega,k) y_i^{(0)}, \ \ \eta_i= \tilde{G}_{ij}(r,\omega,k) \eta_i^{(0)}.
\end{equation}
The parity symmetry of the equations of motion imply that the components $G_{5 i}, G_{i5}, \tilde{G}_{5 i}, \tilde{G}_{i5}$ for $i\neq 5$ are odd in momentum, while the rest of components are even.  Since the elements $\cM_{i5}$, $\cM_{5i}$, $\cN_{i5}$ and $\cN_{5i}$ are all zero for $i\neq 5$,  the current evaluated at the horizon will be even in momentum when the parity odd sources are zero $y_5^{(0)}=\eta_5^{(0)}=0$ or when the parity even sources are zero $y_{i\neq 5}^{(0)}=\eta_{i\neq 5}^{(0)}=0$. In these two cases the current should be invariant under $k\to -k$:
\begin{equation}
J(k)=J(-k).
\end{equation}
However, if both parity even and parity odd sources are nonzero, in general the current will have contributions that are odd in momentum, in contrast to the case of the shear viscosity. We will denote the odd part of the horizon current as
\begin{equation}
[J_H]_{\text{odd}}=\frac{1}{2}\left( J_H(k)-J_H(-k)\right).
\end{equation}

\subsection{Boundary coefficients and correlators}

The asymptotic expansion of metric and scalar  fluctuations takes the form
\begin{equation}
h_{\mu\nu}= e^{2r} \left( h_{\mu\nu}^{(0)}+e^{-3 r} G^T_{\mu\nu}+\cdots\right), \ \ \delta\phi =e^{-(3-\Delta)r} \delta\lambda+e^{-\Delta r} G^O+\cdots.
\end{equation}
We can identify $h_{\mu\nu}^{(0)}$ with a change of the metric in the dual field theory $g_{\mu\nu}^{(0)}=\eta_{\mu\nu}+h_{\mu\nu}^{(0)}$, that acts a as a source for the energy-momentum tensor. Similarly, $\delta\lambda=\mu^{\Delta-3}\delta J^{(0)}$ is a change of the coupling that acts as a source for the scalar operator. The changes in the expectation values of the energy-momentum tensor $\delta\vev{T_{\mu\nu}}$  and scalar $\delta\vev{\cO}$ are proportional to the coefficients $G^T_{\mu\nu}$ and $G^O$ respectively. We have used the holographic renormalization procedure to compute the change in the one-point functions relative to the background values given in \eqref{energy} and \eqref{vevO}
\begin{equation}
\begin{split}
\delta\vev{\cO}=& \frac{\mu^{\Delta-3}}{8\pi G_N L} (3-2\Delta)G^O,\\
\delta\vev{T_{\mu\nu} }=& \frac{3}{16\pi G_N L}G^T_{\mu\nu}+\frac{(\Delta-3)(\Delta-1)}{ 2\Delta-3 }\left[ \vev{\cO}\left(J^{(0)} h_{\mu\nu}^{(0)}+\delta J^{(0)} \eta_{\mu\nu}\right)+J^{(0)}\delta\vev{\cO}\eta_{\mu\nu}\right] .
\end{split}
\end{equation} 
The coefficients $G$ are not independent, but they will be fixed in terms of the sources once regularity or ingoing boundary conditions are imposed on the solutions. In general they will have an expansion
\begin{equation}
\begin{split}
G^T_{\mu\nu}=& G^{TT \ \alpha\beta}_{\mu\nu} h_{\alpha\beta}^{(0)}+G^{TO}_{\mu\nu} \delta \lambda,\\
G^O=& G^{OT\ \alpha\beta} h_{\alpha\beta}^{(0)}+G^{OO} \delta \lambda.
\end{split}
\end{equation}
Where the coefficients $G^{TT}$, $G^{TO}$, $G^{OT}$ and $G^{OO}$ are functions of the frequency and the momentum.

We can derive the correlators of the energy-momentum tensor and scalar by taking variations with respect to the one-point functions
\begin{equation}
\begin{split}
&\Gamma^{TT}_{\mu\nu\alpha\beta}=-\eta_{\alpha\sigma}\eta_{\beta\rho}\frac{\delta \vev{T_{\mu\nu}} }{\delta h_{\sigma\rho}^{(0)}}, \ \ \Gamma^{TO}_{\mu\nu}=\frac{\delta \vev{T_{\mu\nu}} }{\delta J^{(0)}},\\
&\Gamma^{OT}_{\alpha\beta}=-\eta_{\alpha\sigma}\eta_{\beta\rho}\frac{\delta \vev{\cO} }{\delta h_{\sigma\rho}^{(0)}}, \ \  \Gamma^{OO}=\frac{\delta \vev{\cO} }{\delta J^{(0)}}.
\end{split}
\end{equation}

\subsection{Ward identities}

We have now all the ingredients to derive Ward identities. Let us start with the usual Ward identities for the conservation and the trace of the energy-momentum tensor. When we compute the solutions we find that not all the coefficients $G^T_{\mu\nu}$ are independent. They satisfy a linear relation, that in terms of the one-point functions becomes the trace Ward identity
\begin{equation}
\eta^{\mu\nu}\delta\vev{T_{\mu\nu}}-h^{(0)\,\mu\nu}\vev{T_{\mu\nu}}_T = (\Delta-3)\left[ \delta J^{(0)}\vev{\cO}+J^{(0)}\delta\vev{\cO}\right].
\end{equation} 
Where $\vev{T_ {\mu\nu}}_T$ is the thermal energy-momentum tensor determined in \eqref{energy}. 

The momentum constraint equations give two more Ward identities related to the conservation of the energy-momentum tensor
\begin{equation}\label{wardT}
\begin{split}
0=&\omega \delta\vev{T_{00}}+k\delta\vev{T_{10}} +\varepsilon (\omega h_{00}^{(0)}+k h_{01}^{(0)})+\frac{\varepsilon+P}{2}\omega( h_{11}^{(0)}+ h_{22}^{(0)})-\omega \delta J^{(0)} \vev{\cO},\\
0=&\omega \delta\vev{T_{01}}+k\delta\vev{T_{11}} -P (\omega h_{01}^{(0)}+k h_{11}^{(0)})-\frac{\varepsilon+P}{2}k  h_{00}^{(0)}+k\delta J^{(0)}\vev{\cO}.
\end{split}
\end{equation}
These are consistent with the covariant form of the Ward identity expanded to linear order
\begin{equation}
\partial_\mu \left(\delta \vev{T^\mu_{\ \nu}}-h^{(0)\,\mu \alpha}\vev{T_{\alpha \nu}}_T\right)+\Gamma^{(0)\,\mu}_{\mu \alpha}\vev{T^\alpha_{\ \nu}}_T-\Gamma^{(0)\,\alpha}_{\mu\nu }\vev{T^\mu_{\ \alpha}}_T=-\partial_\mu \delta J^{(0)}\vev{\cO}.
\end{equation}

In order to derive a generalized Ward identity for the scalar modes we can use the same argument we used for the shear modes. The current evaluated at the horizon has a contribution $[J_H]_{\text{odd}}$ odd under $k\to -k$. Since the current is constant in the radial direction $J'=0$, the current evaluated at the boundary must have the same property. This gives the conditions
\begin{equation} \label{godd}
\left[ \cG\right]_{\text{odd}}=\frac{\delta^2 }{\delta H^\dagger \delta H}[J_H]_{\text{odd}}.
\end{equation}
Where $\cG$ was defined in \eqref{JB}.  

We use the basis of fluctuations $y_i$ to compute the current, but then we change to the usual basis of metric and scalar fluctuations $h_{\mu\nu}$, $\delta\phi$ to extract $\cG$ and derive the Ward identities. The map between the leading order terms is
\be\label{mapyh}
\begin{split}
& y_1{}^{\text{(0)}} =  \frac{1}{2}\left(h^{(0)}_{11}+h^{(0)}_{22} +h^{(0)}_{00}\right)\,, \ \ 
y_2{}^{\text{(0)}} = \frac{1}{2}\left(h^{(0)}_{11}+h^{(0)}_{22}-h^{(0)}_{00}\right)\,, \ \ 
y_3{}^{(3-\Delta)} =\delta\lambda\,,\\ 
& y_4{}^{\text{(0)}} = \frac{1}{2}\left(h^{(0)}_{11}-h^{(0)}_{22}\right)\,, \ \ 
 y_5{}^{\text{(0)}} =  h^{(0)}_{01}.  
\end{split}
\ee
If we turn on only the parity odd source $y_5{}^{\text{(0)}} $, then $[J_H]_{\text{odd}}=0$ and the Ward identity is simply
\begin{equation}\label{ward01}
\left[\Gamma_{0101}^{TT}-\overline{\Gamma}_{0101}^{TT}\right]_{\text{odd}}=0.
\end{equation}
If we turn on only the parity even sources, we get a quite complicated expression. It becomes somewhat simpler if we impose on the source the tracelessness condition $\eta^{\mu\nu}h_{\mu\nu}^{(0)}=0 \ \Rightarrow y_2{}^{\text{(0)}}=0$ and set the source for the scalar field to zero $y_3{}^{(3-\Delta)} =0$, but it does not lead to any expression that relates to the Ward identity we are interested in.

We have to allow for both parity odd an parity even sources. We find the following condition
\begin{equation}\label{eq:wodd}
\left[\Gamma^{TT}_{1101}+\overline{\Gamma}^{TT}_{0111}\right]_{\text{odd}}=-W_{\text{odd}},
\end{equation}
where the term that appears on the right hand side is schematically
\begin{equation}\label{wodd}
W_{\text{odd}}=-\alpha_J \frac{\delta^2 [J_H]_{\text{odd}}}{\delta \overline{h}_{01}^{(0)}\delta h_{11}^{(0)}}+\lambda \left[ \alpha_{01} \Gamma^{OT}_{01}+\alpha_{00}\Gamma^{OT}_{00}\right]_{\text{odd}}+\alpha k\omega (\omega^2-k^2)^2 .
\end{equation}
$\alpha_{01}$ is a constant and $\alpha_{00}$ depends on the pressure and the expectation value of the scalar operator. The coefficients $\alpha_J$ and $\alpha$ are dimensionful constants determined by the overall factors that appear in the definition of the correlators $\Gamma$ when we compute them using holographic renormalization.

If we multiply by $k$ this equation and use the Ward identity \eqref{ward1} (with the source for the scalar fluctuation set to zero), such that $k \Gamma^{TT}_{1101}= -\omega  \Gamma^{TT}_{0101}+\omega P$, then
\begin{equation}
\left[k \overline{\Gamma}^{TT}_{0111}-\omega \Gamma^{TT}_{0101}\right]_{\text{even}}= -\omega P-k W_{\text{odd}}\,.
\end{equation}
Multiplying by $\omega$ and using \eqref{ward1}, such that $\omega\overline{\Gamma}^{TT}_{0111}=-k\overline{\Gamma}^{TT}_{1111}+k P$, we obtain the expected form of the Ward identity
\begin{equation}\label{wardbulk}
\left[\omega^2 \Gamma^{TT}_{0101}+k^2 \overline{\Gamma}^{TT}_{1111}\right]_{\text{even}}=(\omega^2+k^2)P+k \omega  W_{\text{odd}}\,.
\end{equation}
This establishes a relation between the momentum or thermal conductivity and the bulk viscosity. However, in contrast to the identity for the shear, we do not know how to completely determine the relation without first solving the equations for the fluctuations. 
\section{Discussion}\label{discuss}
In order to derive relations of the form \eqref{wardids} in holography we have constructed a probability current $J$ from linear fluctuations of the metric and a scalar field in an asymptotically AdS spacetime. This current is independent of the radial coordinate and invariant under parity. Using these properties and comparing the value of the current at the AdS boundary and at the horizon, we found the Ward identities \eqref{wardshear} and \eqref{wardbulk}
\begin{equation}
\left[\omega^2 \Gamma_{0202}+k^2 \overline{\Gamma}_{1212}\right]_{\text{even}}=0, \ \
\left[\omega^2 \Gamma^{TT}_{0101}+k^2 \overline{\Gamma}^{TT}_{1111}\right]_{\text{even}}=(\omega^2+k^2)P+k \omega  W_{\text{odd}}\,.
\end{equation}
An expression for $W_{\text{odd}}$ is given in \eqref{wodd}.  In order to derive the second identity we had to introduce auxiliary fields that allowed us to construct a constant probability current. This introduces an ambiguity, we can choose arbitrarily the boundary conditions of the auxiliary fields. We impose the same boundary conditions for the original fluctuations and the auxiliary fields at the AdS boundary, so the current is completely determined by the solutions to the original fluctuations.

We can define the real part of the momentum conductivity $\kappa$ and the shear and bulk viscosities $\eta$, $\zeta$ from the Kubo formulas\footnote{We expand the viscosity tensor as $\eta^{ijkl}=\eta (\delta^{ik}\delta^{jl}+\delta^{il}\delta^{jk}-\delta^{ij}\delta^{kl})+\zeta \delta^{ij}\delta^{kl}$ }
\begin{equation}
\begin{split}
&\kappa_{ij}= -\frac{1}{\omega} \,{\rm Im}\,\Gamma_{0i0j}(\omega,k),\\ 
&\eta=- \frac{1}{\omega} \,{\rm Im}\, \Gamma_{1212}(\omega,k),\\
&\eta+\zeta = -\frac{1}{\omega}  \,{\rm Im}\,\Gamma_{1111}(\omega,k).
\end{split}
\end{equation}
For low momentum $k$, we can expand each of the transport coefficients in powers of $k$
\begin{equation}
\kappa_{ij}\simeq \kappa_{ij}^{(0)}+(k^2\delta^{ij}-k^i k^j) \kappa_T^{(2)}+k^i k^j \kappa^{(2)}_L+\cdots, \ \ \eta=\eta^{(0)}+O(k^2),\ \ \zeta=\zeta^{(0)}+O(k^2).
\end{equation}
We can also expand $W_{\text{odd}}\simeq kW_{\text{odd}}^{(1)}+\cdots$. From the Ward identities we get the relations
\begin{equation}
\kappa_T^{(2)}=\frac{1}{\omega^2}\eta^{(0)},\ \ \kappa_L^{(2)} =\frac{1}{\omega^2}\left(\eta^{(0)}+\zeta^{(0)}  -{\rm Im}\,W_{\text{odd}}^{(1)}\right).
\end{equation}
The first relation between the transverse component of the conductivity and the shear viscosity agree with field theory results. The second relation between the longitudinal conductivity and the bulk viscosity has the right structure, but we do not know from general arguments what is the contribution from $W_{\text{odd}}$. 
In general, $W_{\text{odd}}$ is an asymmetry in the mixed correlators of momentum and stress. From  \eqref{eq:wodd}
\begin{equation}
{\rm Im}\,W_{\text{odd}}=\left[{\rm Im}\,\Gamma_{0111}-{\rm Im}\,\Gamma_{1101}\right]_{\rm odd}.
\end{equation}
A na\"{\i}ve comparison with the Ward identity (2.23) at zero magnetic field in \cite{Hoyos:2015yna} would fix ${\rm Im}\,W_{\text{odd}}^{(1)}=0$. Although this probably holds in the holographic model, the correlators computed using holographic renormalization can differ by contact terms from the correlators that enter in the Ward identity in \cite{Hoyos:2015yna}, so there might be additional contributions. It would be interesting to look for a general argument that fixes the asymmetry in holographic models.

In the calculation using the probability current $W_{\text{odd}}$ contains two kind of contributions, one is coming from the evaluation of the probability current at the boundary and it is ambiguous because the probability current we have constructed depends on auxiliary fields whose boundary conditions can be fixed in different ways. The second kind of contribution depends on the value of the current at the horizon and it cannot be determined without explicitly solving the equations of motion. Since the correlators $\Gamma$ are defined only in terms of the original fluctuations, the horizon and boundary ambiguities should cancel each other, but we cannot determine completely the Ward identity from parity invariance of the current alone. The situation is somewhat improved when only parity even or parity odd sources are turned on, in this case there are no spurious contributions from the horizon. 

Even if we focus on the identity for the transverse component \eqref{wardshear} our derivation of the Ward identity is not complete, it is restricted to terms that are even in momentum in the correlators. In principle we do not expect odd terms appearing in this identity when parity is not broken, but the argument we used for the even terms does not apply to odd terms. This suggests that there must be a  different, more general, derivation of the Ward identities. 

A natural generalization of this work would be to derive similar Ward identities in holographic models with broken parity, in particular the relation between Hall viscosity and Hall conductivity. This is a direction that has not been explored much, even though there are a large variety of models that exhibit a non-zero Hall conductivity: dyonic black holes \cite{Hartnoll:2007ai,Hartnoll:2007ih,Goldstein:2010aw,Gubankova:2010rc,Lippert:2014jma,Lindgren:2015lia}, D-brane intersections of different types \cite{Fujita:2009kw,Bergman:2010gm,Jokela:2011eb,Kristjansen:2012ny,Bea:2014yda} and others \cite{KeskiVakkuri:2008eb,Fujita:2012fp}. However, the value of the Hall viscosity has been determined  in a different class of holographic models dual to parity breaking superfluids \cite{Saremi:2011ab,Son:2013xra,Hoyos:2014nua,Golkar:2015dya}. It would be interesting to check if and when the models that have a Hall conductivity also have a Hall viscosity, since this is mostly the case in Quantum Hall systems and other topological states in condensed matter.

Besides the use we have made of it, the probability current might prove to be useful for other tasks. A possible application is to compute the spectrum of normalizable modes, as are for instance quasinormal modes in a black hole geometry. Let us consider a system with $n$ coupled fluctuations $y_i$, $i=1,\dots,n$ and the related auxiliary fields $\eta_i$. The expansion close to the AdS boundary will include the leading terms of the non-normalizable $y_i^{(d-\Delta_i)}$ (sources) and normalizable $y_i^{(\Delta_i)}$ (vev) solutions of the fluctuations, and similar terms appear in the auxiliary fields (even though the leading terms might be different due to mixing)
\begin{equation}
y_i \simeq y_i^{(d-\Delta_i)}e^{-(d-\Delta_i)r} + y_i^{(\Delta_i)}e^{-\Delta_i r}, \ \ \ \eta_i \simeq \eta_i^{(d-\Delta_i)}e^{-(d-\Delta_i)r} + \eta_i^{(\Delta_i)}e^{-\Delta_i r}+\cdots.
\end{equation}
 If the sources are zero $y_i^{(d-\Delta_i)}=\eta_i^{(d-\Delta_i)}=0$, the  probability current will vanish. This will be independent of the value of the auxiliary fields at the horizon. The solutions for auxiliary fields can be computed by shooting from the AdS boundary with normalizable boundary conditions and do not have to satisfy any regularity conditions at the horizon. We will have $n$ independent solutions that we can construct by imposing $\eta^{(\Delta_i)}=0$ for all modes but one. Then, the condition that the probability current is zero at the horizon for each case will lead to $n$ linear equations for the values of the fluctuations at the horizon $y_i^H$. In order to have a non-trivial solution the system must be degenerate, which will give a condition on the spectrum. This method is somewhat similar to the determinant method of \cite{Amado:2009ts}, but there the system of linear equations is found by evaluating the solutions $y_i$ with ingoing boundary conditions at a cutoff close to the boundary.

\section*{Acknowledgements}
We would like to thank Ioannis Papadimitriou for useful comments. This work is partially supported by the Spanish grant 
MINECO-13-FPA2012-35043-C02-02. C.H. is supported by the Ramon y Cajal fellowship RYC-2012-10370. DRF is supported by the GRUPIN 14-108 research grant from Principado de Asturias.


\appendix

\section{Holographic renormalization}\label{apren}

In order to compute expectation values and correlation functions of operators in the field theory dual, we follow the holographic renormalization prescription  \cite{deHaro:2000xn,Papadimitriou:2004rz}. We will write the metric as
\begin{equation}
ds^2 =dr^2+g_{\mu\nu}dx^\mu dx^\nu.
\end{equation}
The metric asymptotes an $AdS$ space of radius $L$, $g_{\mu\nu}\sim e^{2r/L}$ when $r\to \infty$.
 
We can obtain the one-point functions of the scalar operator and the energy-momentum tensor by taking variation of the action with respect to the metric and the scalar field. Since the action is divergent we need to regularize it, this can be done by introducing a radial cutoff $r_\Lambda$. In order to have a well-defined variation with respect to the metric we have to add a Gibbons-Hawking term at the cutoff 
\begin{equation}
S=\frac{1}{16\pi G_N} \int d^4 x \sqrt{-g}\left( R-(\partial\phi)^2-2 V(\phi)\right)+\frac{1}{8\pi G_N}\int_{r=r_\Lambda} d^3 x \sqrt{-g} K.
\end{equation}
Where $K=g^{\mu\nu}K_{\mu\nu}$ and  $K_{\mu\nu}$ is the extrinsic curvature on radial slices
\begin{equation}
K_{\mu\nu}=\frac{1}{2}\partial_r g_{\mu\nu}.
\end{equation}
The variation of the on-shell bulk action plus Gibbons-Hawking term is
\begin{equation}
\delta S_{\text{on-shell}} =  \frac{1}{16\pi G_N}\int_{r=r_\Lambda} d^3 x\, \sqrt{-g}\left[\left(K_{\mu\nu}-g_{\mu\nu}K \right)\delta g^{\mu\nu}-2\partial_r\phi \delta\phi\right].
\end{equation}
The on-shell action has divergent terms when $r_\Lambda\to \infty$. They can be removed by adding a counterterm action at the cutoff
\begin{equation}
S_{c.t.}=\frac{L}{8\pi G_N}\int_{r=r_\Lambda} d^3 x\, \sqrt{-g}\left( -\frac{2}{L^2}+\frac{\Delta-3}{2L^2}\phi^2-\frac{1}{2}\hat{R}\right),
\end{equation}
where $\hat{R}$ is the Ricci scalar of the induced metric on the radial slice. The variation of the sources for the dual operators $ \delta g^{(0)\, \mu\nu}$, $ \delta J^{(0)}$ are identified as
\begin{equation}
\delta g^{\mu\nu}=e^{-2 r/L} \delta g^{(0)\, \mu\nu},\ \  \delta \phi = \mu^{\Delta-3} \delta J^{(0)}e^{-(3-\Delta)r/L}.
\end{equation}
The one-point functions are computed from the variation of the action with respect to the sources
\begin{equation}
\vev{T_{\mu\nu}}=-\lim_{r_\Lambda\to \infty} \frac{2}{\sqrt{-g^{(0)}}}\frac{\delta S}{\delta g^{(0)\,\mu\nu}},\ \ \vev{\cO}=-\lim_{r_\Lambda\to\infty} \frac{1}{\sqrt{-g^{(0)}}}\frac{\delta S}{\delta J^{(0)}}.
\end{equation}

The finite one-point functions are defined as
\begin{equation}
\begin{split}
\vev{T_{\mu\nu}}=&  -\lim_{r_\Lambda \to \infty} \frac{1}{8\pi G_N}e^{-2r_\lambda/L}\frac{\sqrt{-g}}{\sqrt{-g^{(0)}}} \times \nonumber\\
&\left[  K_{\mu\nu}-g_{\mu\nu}K +\frac{1}{L}g_{\mu\nu}\left( 2-\frac{ \Delta-3}{2}\phi^2\right)-L\left(\hat{R}_{\mu\nu}-\frac{1}{2}g_{\mu\nu}\hat{R}\right)\right]_{r=r_\Lambda},\\
\vev{\cO}=& \lim_{r_\Lambda\to \infty} \frac{\mu^{\Delta-3}}{8\pi G_N}e^{-(3-\Delta)r_\Lambda/L}\frac{\sqrt{-g}}{\sqrt{-g^{(0)}}}\left[ \partial_r \phi -\frac{\Delta-3}{L} \phi\right]_{r=r_\Lambda}.
\end{split}
\end{equation}

\section{Equations of motion}\label{app:eqs}

In this appendix, we write down the equations of motion derived from the action \Eq{classac}. In principle, we are not interested on the full solution of neither the metric functions $A\,,B$, nor of the scalar field $\phi$, since the form of the asymptotic expansions suffices in this work. The same applies to the fluctuations $\lbrace y_i\rbrace,\lbrace \eta_i \rbrace \,, i= 1,\cdots,5$.\\

\subsection{Background equations}\label{app:backeqs}

For the black brane background \Eq{gbulk} plus scalar field $\phi=\phi_0(r)$ coupled to gravity, in absence of fluctuations the background equations of motion read (the $r$ dependence is implicit)
\bea \label{eoms0}
0 &=& \phi _0' \left[3A'+B'\right]-V'\left(\phi
_0\right)+\phi _0''\,,\\
0 &=& A''-A' B'+\frac{1}{2} \phi _0'{}^2\,,\\
0&= & B' \left[3 A'+B'\right]+B''\,,\\
V\left(\phi _0\right) &=& \frac{1}{2}\phi_0'{}^2 -
A' \left[2B' +3 A'\right]\label{V}\,,\\
\nonumber
\eea
where the constraint was employed to set the potential as a function of the derivatives of $A\,,B$ and $\phi_0$ alone\footnote{The prime will denote derivative with respect to the radial coordinate, except for the potential $V(\phi_0)$ where it denotes derivative with respect to the field $\phi_0$.}. We have freedom to choose the potential $V$, for the sake of simplicity we will restrict ourselves to a quadratic potential on the field $\phi$, i.e., $V(\phi)\propto \phi^2$. This choice does not affect the results but simplifies somewhat the formulas.\\

\subsection{Fluctuation equations}\label{app:fluceqs}

The equations of fluctuations are
\begin{equation}
y_i''+a_{ij}\, y_j'+b_{ij}\, y_j=0, \ \ c_i^a\,y_i'+d_i^a\, y_i=0, \ a=1,2,3.
\end{equation}
Where the coefficients of the dynamical equations are
\begin{equation}
\arraycolsep=6pt
a_{ij}=\left(
\begin{array}{ccccc}
B' & e^{-\frac{3 A}{2}} \left(A'-B'\right) &  &  &  \\
& 2 B' &  &  &  \\
& e^{-\frac{3 A}{2}} \phi_0' & B' &  & \\
&  &  & B' &  \\
&  &  &  & 3 B' 
\end{array}
\right),
\end{equation}

\begin{align}
b_{12} &= - e^{-\frac{7}{2}A} \left[\frac{k^2}{2} +3 e^{2 A} A'   \left(A'-B'\right)\right]\,, & b_{13} &= V'\left(\phi _0\right)\,,\nonumber \\
b_{14} &= e^{-2 A} k^2 \,, & b_{21} &=  e^{-\frac{A}{2}} \left( \frac{k^2}{2}+ e^{-2 B}\omega ^2\right)\,, \nonumber \\
b_{23} &= 3 e^{\frac{3 A}{2}} V'\left(\phi _0\right)\,, & b_{24} &=  e^{-\frac{A}{2}} k^2\,,\nonumber\\
b_{25} &= 2 e^{-\frac{A}{2}} k \omega\,,& b_{32} &= -3 e^{-\frac{3 A}{2}} A' \phi _0'\,,\nonumber\\
b_{41} &= \frac{k^2}{2} e^{-2 A}  \,, & b_{42} &= -\frac{k^2}{2} e^{-\frac{7 A}{2}} \,,\nonumber\\
b_{45} &= e^{-2 A} k \omega\,,& b_{51} &= \frac{1}{2} e^{-2 (A+B)} k \omega\,,\nonumber\\
b_{52} &= \frac{1}{2} e^{-\frac{7 A}{2}-2 B} k \omega\,, & b_{54} &= -e^{-2 (A+B)} k \omega\,,\nonumber
\end{align} 
\bea 
b_{11} &=& - \frac{1}{2} e^{-2 A} k^2+ \frac{3}{4} \left[\phi _0'{}^2-  A' \left(3 A'+4 B'\right)\right]\,,\nonumber\\
b_{22} &=& \frac{3}{2} \phi _0'{}^2+ e^{-2A} \left( e^{-2 B}\omega ^2-  \frac{3k^2}{2}\right)-9 A' \left(A'+B'\right)\,,\nonumber \\
b_{33} &=& -3A'\left(B'+\frac{3}{4}A' \right)+\frac{3}{4} \phi _0'{}^2+e^{-2A} \left(e^{-2 B}\omega ^2- k^2\right)-V''\left(\phi _0\right)\,,\nonumber\\
b_{44} &=& e^{-2 (A+B)} \omega ^2+\frac{3}{4} \phi _0'{}^2-3A'\left(B'+\frac{3}{4}A' \right)\,,\nonumber\\
b_{55} &=& \frac{3}{4} \left[\phi _0'{}^2-A'\left(3 A' +8 B'\right)\right]\,.\nonumber
\eea

The coefficients of the constraints are
\begin{equation}
c^1_i = -\frac{\w}{2} \left(\begin{array}{c} 1 \\ 1 \\ 0 \\ 0 \\ \frac{k}{\w}  e^{2 B(r)} \end{array}\right),
\ c^2_i =  -\frac{k}{4}\left(\begin{array}{c} 1 \\ -3 \\ 0 \\ 2 \\ \frac{2\w}{k} \end{array}\right),
\ c^3_i= \left(\begin{array}{c} \frac{B'}{2} \\ \frac{1}{2} \left(4 A'+ B'\right) \\ -\phi_0' \\ 0 \\ 0 \end{array}\right)\,,
\end{equation}
and  
\begin{equation}
d^1_i = \frac{\w}{2} \left(\begin{array}{c}  B'(r)\\  B'(r) \\ -2 \w \phi_0' \\ 0 \\ 0 \end{array} \right),
\ d^2_i = \frac{k}{2} \left(\begin{array}{c} - B' \\  B' \\ 2 \phi_0' \\ 0 \\ -\frac{2\w}{k} B' \end{array}\right),
\ d^3_i= 4e^{-2A}\left(\begin{array}{c} 2 \omega ^2e^{-2 B} +  k^2 \\ 2 \omega ^2e^{-2B} -3 k^2   \\ 4e^{2A} V'(\phi) \\ 2 k^2
\\ k \omega   \end{array}\right).
\end{equation}

\subsection{Coefficients in the boundary  current}\label{app:currcoefs}

The current evaluated at the boundary has the form
\begin{equation}
J_B= H^\dagger \cC H + H^\dagger \cD T -T^\dagger \cD^\dagger H,
\end{equation}
where $\cC^\dagger = -\cC$. The non-zero coefficients of $\cC$ are
\begin{equation}\
\begin{split}
&\cC_{12}=-\cC_{21}=\frac{1}{64}(k^2+2\omega^2)(\omega^2-k^2)^2+\frac{13}{6} B_0-\frac{1}{2}\Delta(\Delta-3) \lambda v,\\
&\cC_{13}=-\cC_{31}=-\frac{1}{3}\Delta(\Delta-3) v,\\
&\cC_{23}=-\cC_{32}=-\frac{\Delta(2\Delta-9)}{2\Delta+3}v+\frac{3}{4}\Delta(2\Delta-7) B_0 v-\frac{1}{32}(\Delta(2\Delta-3)^2-81)\lambda v^2,\\
&\cC_{24}=-\cC_{42}= -\frac{1}{32}k^2(\omega^2-k^2)^2,\\
&\cC_{25}=-\cC_{52} = -\frac{1}{16}k\omega (\omega^2-k^2)^2.
\end{split}
\end{equation}
The non-zero components of $\cD$ are
\begin{equation}
\begin{split}
& 2\cD_{11}=\cD_{22}=2\cD_{44}=2\cD_{55}=-4\cD_{21}= 12,\\
&\cD_{33}=2(2\Delta-3),\\
&\cD_{13}=-\frac{1}{3}\Delta(\Delta-3) \lambda,\\
&\cD_{23}=\frac{3}{4}(\Delta-3)(2\Delta+1) \lambda B_0+\frac{(\Delta-3)(2\Delta+2)}{2\Delta-9} \lambda 
+\frac{1}{32}\left( \Delta(2\Delta-3)(2\Delta-9)+18(\Delta+3)\right)\lambda^2 v.
\end{split}
\end{equation}

\subsection{Coefficients in constraints}\label{app:constrcoefs}

The components of $\cC^a$ and $\cD^a$ are
\begin{equation}
\begin{split}
&\cC_i^1 =-\w\left\{\frac{1}{6}  \left[3 B_0+(\Delta -3) \Delta  \lambda  v\right],B_0
,-\frac{ \Delta  v }{3} (\Delta -1) ,0,\frac{k}{6\w}  \left[(\Delta -3) \Delta  \lambda  v-3
B_0\right]\right\},\\
&\cD_i^1 = \left\{\omega ,0,\frac{1}{3} (\Delta -3) (\Delta -2) \lambda  \omega ,0,k\right\}, \\
&\cC_i^2 =-k\Bigg\lbrace\frac{1}{12}  \left[(\Delta -3) \Delta  \lambda  v-15 B_0\right],B_0
,\frac{1}{6} \Delta  (3 \Delta -5) v,\nonumber\\
& \frac{1}{6} \left[(\Delta
-3) \Delta  \lambda  v -3 B_0\right],\frac{\omega}{6 k}   \left[(\Delta -3) \Delta 
\lambda  v-15 B_0\right]\Bigg\rbrace,\\
&\cD_i^2 =\left\{\frac{k}{2},0,-\frac{1}{6} (\Delta -3) (3 \Delta -4) k \lambda ,k,\omega
\right\}, \\
& \cC^3_1 = B_0^2 \left(k^2+\omega ^2\right)\,,\quad \cC^3_4 = \frac{1}{6} B_0 k^2 \left(3 B_0-(\Delta -3) \Delta  \lambda  v\right)\,,\quad \cC^3_5 = 2 B_0^2 k \omega\,,\\
&\cC_2^3 = \frac{B_0}{2
	(\Delta -3) \Delta  \lambda  v \left(k^2-2 \omega ^2\right)}\left[2+\frac{3 B_0 \left(5 \omega ^2-4 k^2\right)}{(\Delta
		-3) \Delta  \lambda  v \left(k^2-2 \omega ^2\right)}-\frac{(2 (\Delta
		-3) \Delta +9) (8 (\Delta -3) \Delta -9) \lambda  v}{48 B_0 (\Delta -3) \Delta }\right]\,,\\
&\cC_3^3 = -\frac{v}{3}\Bigg\lbrace \frac{2 B_0 \Delta \left[(2 \Delta -5) \omega ^2+k^2\right]}{ \left(k^2-2
	\omega ^2\right)} + \frac{1}{72} (2 (\Delta -3) \Delta +9) (8 (\Delta -3) \Delta
-9) \lambda  v\Bigg\rbrace\,,\\ 
& \cD_{1,5}^3 = 0\,,\quad \cD_2^3 = \frac{4}{3} \left(k^2-2 \omega ^2\right)\,,\quad \cD_4^3 = B_0 k^2\,,\\
&\cD_3^3 = \frac{1}{3}(\Delta -3) \lambda  \left(k^2-2 \omega ^2\right)\Bigg\lbrace\frac{2 B_0 \left[(1-2 \Delta ) \omega ^2+k^2\right]}{k^2-2 \omega ^2}-\frac{(2
	(\Delta -3) \Delta +9) (8 (\Delta -3) \Delta -9) \lambda  v}{72 (\Delta -3)}\Bigg\rbrace.\\
\end{split}
\end{equation}

\section{Series expansions}\label{app:exp}
In this appendix, we will detail the form of the on-shell series expansions which have been used in this work, both for the background functions and for the (original-auxiliary) fluctuations. \\
	
\subsection{Background at the boundary}\label{app:backexpb}
	
Since AdS is an asymptotic fixed point when $r\ra \infty$, we must impose that at leading order $A\sim r\,,B\sim 0\,,\phi_0\sim 0$ at the boundary. We express the subleading contribution as the sums
\bea 
\widetilde{A}(r) &\sim & \sum_{n,m} a_{(n,m)} e^{-(n + m\Delta) r}\,, \label{ab}\\
B(r) &\sim &\sum_{n,m} b_{(n,m)} e^{-(n -\Delta) r}\,, \label{bb}\\
\phi_0(r) &\sim & \sum_{n,m} \phi_{(n,m)}e^{-(3n +\Delta m) r} \label{fb}\\
\nonumber
\eea 
where $A(r) = r +\widetilde{A}(r)$. $a,b,\phi_{(n,m)}$ are real-valued coefficients and $\phi_{(1,-1)}=\lambda,\,\phi_{(0,1)}=v,\,b_{(1,0)} = B_0$.  Combining \Eq{ab}-\Eq{fb} with \Eq{eoms0}, 
\bea
\widetilde{A}(r) &=& a_{(2,-2)} e^{-2(3 r - \Delta)  r} +  a_{(1,0)}e^{-3 r} + a_{(0,2)} e^{-2 \Delta  r} + \cdots \,,\\
B(r) &=& B_0 e^{-3 r} + b_{(3,-2)} e^{-(9- 2 \Delta ) r} +  b_{(2,0)}e^{-6 r} + b_{(1,2)} e^{-(3 + 2 \Delta)  r} + \cdots \,,\\
\phi_0(r) &=& \lambda  e^{-(3  -\Delta)  r} + v e^{-\Delta r} + \phi_{(3,-3)} e^{-3(3 - \Delta)  r} +   \phi_{(2,-1)} e^{-(6 - \Delta)  r} + \nonumber\\
& & \phi_{(1,1)} e^{-(3 + \Delta)  r} + \phi_{(0,3)} e^{-3 \Delta  r} + \cdots \,,\\
\nonumber
\eea 
\begin{align}
a_{(2,-2)} &=-\frac{ \lambda ^2}{8} ,\,\, & a_{(1,0)} &= \frac{1}{9} \left[(\Delta -3) \Delta  \lambda  v-3 B_0\right]\,, \nonumber\\
a_{(0,2)} &= -\frac{v^2}{8}\,, & b_{(3,-2)} & = \frac{9 B_0 \lambda ^2}{72-16 \Delta }\,, \nonumber\\
b_{(2,0)} &= -\frac{1}{6} B_0 (\Delta -3) \Delta  \lambda v,\,\, & b_{(1,2)} &= \frac{9 B_0 v^2}{8 (2 \Delta +3)},\,\,\nonumber\\
\phi_{(3,-3)} &= \frac{3 (\Delta -3) \lambda ^3 }{8 (4 \Delta -9)},\,\, &\phi_{(2,-1)} &= \frac{\Delta  (4 \Delta -15) \lambda ^2 v}{24 }\,,\nonumber\\
\phi_{(1,1)} &= -\frac{\lambda  v^2}{24} (\Delta -3) (4 \Delta +3) ,\,\, &\phi_{(0,3)} &= \frac{3 \Delta  v^3}{8 (4 \Delta -3)}\,,
\end{align}
with $V(\phi) \sim \Delta(\Delta -3) \phi^2$. \\

\subsection{Background at the horizon}\label{app:backexph}

As it was stated in Sec.\ref{sec:hocu}, we impose regularity of each of the background functions $A\,,\phi_0$ but $B$, which diverges logarithmically at the horizon 
\be
A(r) = \sum_{n=0} A_H^{(n)}(r-r_H)^n\,, \quad  B(r) = \log (r-r_H) + \sum_{n=0} B_H^{(n)}(r-r_H)^n \,, \quad \phi_0(r) = \sum_{n=0} \phi_{H}^{(n)}(r-r_H)^n\,.
\ee

Plugging these expansions into the equations of motion, 
\bea 
A(r) &=& A_H +\left(r-r_H\right)^2 A_H^{\text{(2)}} \left(r-r_H\right)^4 A^{(4)}_H + (r-r_H)^6 A^{(6)}_H +\cdots \,,\\
B(r) &=& \log (r-r_H) + B_H + \left(r-r_H\right)^2 B^{(2)}_H +\left(r-r_H\right)^4 B_H^{(4)}+ \nonumber \\
& & \left(r-r_H\right)^6 B^{(6)}_H + \cdots \,,\\
\phi_0(r) &=&  \phi_H +\left(r-r_H\right)^2 \phi^{(2)}_H + \left(r-r_H\right)^4 \phi^{(4)}_H + \left(r-r_H\right){}^6 \phi^{(6)}_H + \cdots \,,
\eea
up to $\mathcal{O}(r-r_H)^7$, $A_H^{\text{(0)}}=A_H,\,B_H^{\text{(0)}}=B_H,\,\phi^{(0)}_H= \phi_H$ in Eq.\Eq{backhor} and
\begin{align}
A^{(4)} &=-\left(\frac{1}{2}A_H^{\text{(2)}}{}^2 +\frac{1}{64} V'\left(\phi_H\right)^2\right)\,, & A^{(6)}_H &=\frac{3}{160} V'\left(\phi_H\right){}^2
A_H^{\text{(2)}}+\frac{2}{5} A_H^{\text{(2)}}{}^3-\frac{1}{768} V'\left(\phi_H\right)^2 V''\left(\phi_H\right)\,,\nonumber\\
B^{(2)}_H &= -A_H^{\text{(2)}}\,, & B^{(4)}_H &= \frac{7}{10}A_H^{\text{(2)}}{}^2+\frac{3}{320} V'\left(\phi_H\right)^2\,,\nonumber\\
\phi^{(2)}_H &= \frac{1}{4}  V'\left(\phi_H\right)\,, & \phi^{(4)}_H &= \frac{1}{64}  V'\left(\phi_H\right)\left( V''\left(\phi_H\right)-8 A_H^{\text{(2)}}\right) \nonumber\,,
\end{align}
\bea
B^{(6)}_H &=& -\frac{3}{160} V'\left(\phi_H\right)^2
A_H{}^{\text{(2)}}-\frac{62}{105} A_H^{\text{(2)}}{}^3+\frac{V'\left(\phi
	_H\right){}^2 V''\left(\phi_H\right)}{1792}\,,\nonumber\\ 
\phi^{(6)}_H &=& V'\left(\phi
_H\right) \left( \frac{1}{10} A_H{}^{\text{(2)}}{}^2-\frac{1}{96}  V''\left(\phi_H\right) A_H^{\text{(2)}}+\frac{1}{480} V'\left(\phi
_H\right){}^2+\frac{ V''\left(\phi_H\right){}^2}{2304}\right)\,.\nonumber\\
\eea

\subsection{Matrix K}\label{app:Keqs}

Near the horizon, since $e^{B}\sim \left(r-r_H \right)$, we take
\begin{align}
K_{11,33,44} &=  \sum_{n=0} \widehat{K}^{(n)}_{11,33,44}(r-r_H)^{2n +1},\,\,& K_{22} &= \sum_{n=0} \widehat{K}^{(n)}_{22} (r-r_H)^{2n+2},\,\,\nonumber\\
K_{55} &= \sum_{n=0} \widehat{K}^{(n)}_{55} (r-r_H)^{2n+3}\,, & K_{12} &= \sum_{n=0} \widehat{K}^{(n)}_{12} (r-r_H)^{2n+1} + \widehat{K}^{(n,1)}_{12} (r-r_H)^{2n+2}\,,\nonumber
\end{align}
\be 
K^H_{32} = \sum_{n=0} \widehat{K}^{(n)}_{32} (r-r_H)^{2n+2} + \widehat{K}^{(n,1)}_{32} (r-r_H)^{2n +3}\,,\nonumber
\ee

Truncating these expansions at fair enough order, 
\bea 
K_{11} &\sim& \left(r-r_H\right)
K^{H}_{11} + \left(r-r_H\right)^3 \widehat{K}_{11}^{\text{(1)}}
+ (r-r_H)^5\widehat{K}^{(2)}_{11} +  \cdots \,,\nonumber\\
K_{12} &\sim& \left(r-r_H\right)K^{H}_{12} +\left(r-r_H\right)^2
\widehat{K}_{12}^{\text{(0,1)}}  + \left(r-r_H\right)^3 \widehat{K}^{(1)}_{12} + \left(r-r_H\right)^4 \widehat{K}^{(1,1)}_{12} + (r-r_H)^6\widehat{K}^{(2,1)}_{12} \cdots\,,\nonumber\\
K_{22} &\sim& \left(r-r_H\right)^2 K^{H}_{22} + 
\left(r-r_H\right)^4 \widehat{K}_{22}^{\text{(1)}}+ (r-r_H)^6\widehat{K}^{(2)}_{22} + \cdots \,,\nonumber\\
K_{32} &\sim&  \left(r-r_H\right)^2 K^{H}_{32} + \left(r-r_H\right)^3\widehat{K}^{(0,1)}_{32}+\left(r-r_H\right)^4 \widehat{K}_{32}^{\text{(1)}} +(r-r_H)^5\widehat{K}^{(1,1)}_{32}+ (r-r_H)^6\widehat{K}^{(2)}_{32} \cdots \,,\nonumber\\
K_{33} &\sim& \left(r-r_H\right)
K^{H}_{33} + \left(r-r_H\right)^3
\widehat{K}_{33}^{\text{(1)}}  + (r-r_H)^5 \widehat{K}^{(2)}_{33} + \cdots \,,\nonumber\\
K_{44} &\sim&  \left(r-r_H\right)
K^{H}_{44} + \left(r-r_H\right){}^3
\widehat{K}_{44}^{\text{(1)}}  + (r-r_H)^5 \widehat{K}^{(2)}_{44} + \cdots \,, \nonumber\\
K_{55} &\sim&  \left(r-r_H\right)^3 K^{H}_{55} +  (r-r_H)^5 \widehat{K}^{(2)}_{55} + \cdots \,,\nonumber
\eea 
with 
\begin{align}
\widehat{K}_{11}^{\text{(1)}} &= -A_H^{\text{(1)}}K^H_{11}\,, &\widehat{K}^{(1,1)}_{12} &= -2A_H^{\text{(1)}}\widehat{K}_{12}^{\text{(0,1)}}\,,\nonumber\\
 \widehat{K}^{(1)}_{12}& =\frac{5}{2}K^H_{11}e^{-\frac{3}{2}A_H^{\text{(0)}}}A_H^{\text{(1)}}\,, &\widehat{K}^{(0,1)}_{12} &=  e^{-\frac{3}{2}A_H^{\text{(0)}}}K^H_{11}\,,\nonumber\\
\widehat{K}_{22}^{\text{(1)}} &= -2 A_H^{\text{(1)}}K^H_{22}\,, &\widehat{K}_{32}^{\text{(1)}} &= -2A_H^{\text{(1)}}K^H_{32}\,, \nonumber\\
\widehat{K}^{(0,1)}_{32} &= \frac{1}{2}K^H_{33} V'(\phi_H) e^{-\frac{3}{2} A_H^{\text{(0)}}},\,\,&\widehat{K}^{(1)}_{33,44} &=-A_H^{\text{(1)}}K^H_{33,44}\,,\nonumber\\
 \widehat{K}^{(1,1)}_{32} &= \frac{K^H_{32}}{48}V'\left(\phi _H\right) e^{-\frac{3 A_H^{\text{(1)}}}{2}}
 \left(V''\left(\phi _H\right)-60 A_H^{\text{(2)}}\right)\,, &  \widehat{K}^{(2)}_{55} & = -3 A_H^{\text{(1)}} K^H_{55}\,,\nonumber
\end{align}
\bea
 \frac{\widehat{K}^{(2)}_{22}}{K^H_{22}} &=& \frac{\widehat{K}^{(2,1)}_{12}}{K^H_{12}} = \frac{\widehat{K}^{(2)}_{32}}{K^H_{32}} = \frac{1}{160} \left[544 A^{(1)}_H{}^2+3 V'(\phi_H)^2\right]\,, \nonumber\\
\frac{\widehat{K}^{(2)}_{ii}}{K^H_{ii}} &=& \frac{3}{320}  \left[128A_H^{\text{(2)}}{}^2+V'\left(\phi_H\right)^2\right]\,,\quad i=1,3,4\,. \nonumber\\
\eea
up to $\mathcal{O}(r-r_H)^7$. On the other hand, near the boundary,
\be
K^B_{ij} = \delta_{ij} + \sum_{n,m}\widetilde{K}^{(n,m)}_{ij}e^{-(3 n +\Delta m)r}\,,\quad K^B_{12,32} = e^{-\frac{3r}{2}}\sum_{n,m}\widetilde{K}^{(n,m)}_{12,32}e^{-(3 n +\Delta m)r}\,.
\ee

The on-shell series expansions read
\bea
K^B_{11} &=& 1+ e^{-3 r} \widetilde{K}^{\text{(1,0)}}_{11} + e^{-(9 r - 2 \Delta ) r} \widetilde{K}^{\text{(3,-2)}}_{11} + e^{-6 r} \widetilde{K}^{\text{(2,0)}}_{11} + e^{-(3 r -2 \Delta)  r} \widetilde{K}^{\text{(1,2)}}_{11}+\cdots\,, \nonumber\\
K^B_{12} &=& e^{-\frac{3r}{2}}\left[ \widetilde{K}^{\text{(0,0)}}_{12} + e^{-2(3- \Delta)  r} \widetilde{K}^{\text{(2,-2)}}_{12} + e^{-3r} \widetilde{K}^{\text{(1,0)}}_{12} + e^{-2 \Delta  r} \widetilde{K}^{\text{(0,2)}}_{12} +\cdots\right]\,, \nonumber\\
K^B_{22} &=& 1 + e^{-3 r} \widetilde{K}^{\text{(1,0)}}_{22} + \cdots \,,\nonumber\\
K^B_{32} &=& e^{-\frac{3r}{2}}\left[e^{-(3-\Delta )  r} \widetilde{K}^{\text{(1,-1)}}_{32} + e^{-\Delta  r} \widetilde{K}^{\text{(0,1)} }_{32} + \cdots \right]\,,\nonumber\\
K^B_{ii} &=& 1 + e^{-3 r} \widetilde{K}^{\text{(1,0)}}_{ii}+\cdots \,,\quad i= 3,4,5 \nonumber
\eea
with 
\begin{align}
\widetilde{K}^{\text{(1,0)} }_{11} &= B_0\,, &\widetilde{K}^{\text{(3,-2)}}_{11} &= \frac{9 B_0 \lambda ^2}{8(9-2 \Delta) } \,,&  \widetilde{K}^{\text{(2,0)}}_{11} &= \frac{1}{6} B_0 \left[3 B_0-(\Delta -3) \Delta  \lambda 
v\right]\,, \nonumber\\
\widetilde{K}^{\text{(1,2)}}_{11} &= \frac{9 B_0 v^2}{8 (2 \Delta +3)}\,,
&\widetilde{K}^{\text{(0,0)}}_{11} &= -\frac{2}{3}\,,&\widetilde{K}^{\text{(2,-2)}}_{12} &= -\frac{\lambda ^2}{8}\,,\nonumber\\
\widetilde{K}^{\text{(1,0)}}_{12} &=  \frac{1}{9} \left[(\Delta -3) \Delta \lambda  v-19 B_0\right]\,, &\widetilde{K}^{\text{(0,2)}}_{12} &= -\frac{v^2}{8}\,,
&\widetilde{K}^{\text{(1,0)}}_{22} &= 2B_0\,,\nonumber\\
\widetilde{K}^{\text{(1,-1)}}_{32} &= \frac{2 (\Delta -3) \lambda }{2 \Delta -9}\,, & \widetilde{K}^{\text{(0,1)}}_{32} &= \frac{2 \Delta  v}{2 \Delta +3}\,,&\widetilde{K}^{\text{(1,0)}}_{33} &= \widetilde{K}^{\text{(1,0)}}_{44} = \frac{\widetilde{K}^{\text{(1,0)}}_{55}}{3} = B_0\,,\nonumber\\
\end{align}
plus higher order terms. \\

\subsection{Fluctuations at the boundary}\label{app:flucexpb}
	
The near boundary series expansion of the fluctuations is proposed analogously as for the background functions. However, in this case, in order to capture all possible contributions from backreaction and gravity, we will assume a more complex series expansion than the one considered for a CFT. This time,
\bea
y_{1,4,5} (r) &=&  e^{\frac{3r}{2}}  \sum_{n,m>0} \left( \sum_{l=1}^{4} y_{1,4,5}^{(n,m,-l)}e^{-l(3 - \Delta)r} +\sum_{l=0}^{4}  y_{1,4,5}^{(n,m,l)} e^{-l\Delta r}\right)  e^{-\left(2n + \frac{3 }{2} m \right)r}\,,\\
y_2(r) &=& e^{3r}  \sum_{n,m>0} \left( y_2^{(n,m,-3)}e^{-3(2 - \Delta) r}+ \sum_{l=1,2,4} y_2^{(n,m,-l)}e^{-l(3 - \Delta)r} +\sum_{l=0}^{4} y_2^{(n,m,l)} e^{-l\Delta r}\right) e^{-\left(2n + \frac{3 }{2} m \right)r},\nonumber\\
\\
y_3(r) &=&  e^{-\frac{3r}{2}}  \sum_{n,m>0} \left( y_3^{(n,m,0)} +  e^{3 r}\sum_{l=1}^{4} y_3^{(n,m,-l)}e^{-l(3 - \Delta)r} + y_3^{(n,m,l)}e^{-l\Delta r}\right) e^{-\left(2n + \frac{3 }{2} m \right)r},\nonumber\\
\eea
where the exponential pre-factors are due to the changes of variables \Eq{changes}. The non-normalizable modes are identified as $\lbrace y_3^{(0,0,-1)},y_{1,2,4,5}^{(0,0,0)} \rbrace$, whereas the normalizable as $\lbrace y_{1,4,5}^{(0,2,0)},$ $y_2^{(0,4,0)},y_3^{(0,0,1)}\rbrace $. Up to the non-normalizable mode for each fluctuation, 
\bea
y_1 &=& e^{\frac{3}{2}r}\Big\lbrace  y_1^{(0)} +y_1^{(1)} e^{-2(3-\Delta )r }+ y_1^{(2)}e^{-2r} + y_1^{(3)}e^{-3r} + \cdots\Big\rbrace \,, \nonumber\\
y_2 &=&  e^{3r}\Big\lbrace  y_2^{(0)} +y_2^{(2)}e^{-2r} +  y_2^{(\Delta)} e^{-2(3-\Delta) r} + y_2^{(3)}e^{-3r} + y_2^{(2\Delta)} e^{-2(4-\Delta) r}+y_2^{(3\Delta)}e^{-2(1+\Delta)r} +\nonumber\\
& &  y_2^{(5)}e^{-5 r}+ y_2^{(4\Delta)} e^{-4(3-\Delta)r} + y_2^{(5\Delta)} e^{-(9-2\Delta )r} +  y_2^{(6)}e^{-6 r}  +\cdots \Big\rbrace \,,\\
y_3 &=& e^{-\frac{3}{2}r}\Big\lbrace y_3^{(3-\Delta)} e^{ \Delta  r} + y_3^{(\Delta)} e^{-(\Delta -3)r} +\cdots \Big\rbrace \,,\nonumber\\
y_{4,5} &=&  e^{\frac{3}{2}r}\Big\lbrace y_{4,5}^{(0)} +  y_{4,5}^{(1)}e^{-2r} + y_{4,5}^{(2)} e^{-2(3-\Delta )r} +  y_{4,5}^{(3)}e^{-3 r}+ \cdots\Big\rbrace  \,,\nonumber
\nonumber
\eea
where we have replaced the $(n,m,l)$ numeration by other suited to the one employed along the work
\bea
y_1^{(1)} &=& -\frac{\lambda}{16}   \left(3 \lambda  y_1^{(0)}+4
   y_3^{(3-\Delta)}\right),\,\,\nonumber\\
y^{(2)}_1 &=& -\frac{1}{16} \left[4 k \omega  y_5^{(0)}+k^2 \left(5
y_1^{(0)}-6 y_4^{(0)}+y_2^{(0)}\right)+2
\omega ^2 \left(y_1^{(0)}+y_2^{(0)}\right)\right]\,,\nonumber\\
y^{(2)}_2 &=& \frac{1}{16} \left[4 k \omega  y_5^{(0)}+k^2
   \left(y_1^{(0)}+2 y_4^{(0)}-3
   y_2^{(0)}\right)+2 \omega ^2
   \left(y_1^{(0)}+y_2^{(0)}\right)\right],\nonumber\\
y^{(\Delta)}_2 &=& -\frac{3\lambda}{8}   \left(2 y_3^{(3-\Delta)}+\lambda 
      y_2^{(0)}\right),\,\, \nonumber\\
y_2^{(3)} &=& \frac{1}{3} \Big\lbrace (\Delta -3) \Delta  \left[v
         \left(y_3^{(3-\Delta)}+\lambda  y_2^{(0)}\right)+\lambda 
         y_3^{(\Delta)}\right]-3 B_0 y_2^{(0)}\Big\rbrace \,, \nonumber\\
y_2^{(2\Delta)} &=& \frac{( 10-3 \Delta) \lambda}{128 (\Delta -4) (2 \Delta -5)}  \Big\lbrace 4 \lambda  k \omega y_5^{(0)}+k^2 \left[\lambda  \left(y_1^{(0)}+2y_4^{(0)}-3 y_2^{(0)}\right)-16y_3^{(3-\Delta)}\right]+\nonumber\\
& & 2 \omega ^2 \left[\lambda \left(y_1^{(0)}+y_2^{(0)}\right)+8y_3^{(3-\Delta)}\right]\Big\rbrace\,,\nonumber\\
y_2^{(5)} &=& \frac{-1}{720 \mathcal{G}_1}\Bigg\lbrace -3 B_0 \mathcal{G}_1 \Big[k^2 \left(13 y_1^{(0)}-3 y_2^{(0)}+26 y_4^{(0)}\right)+52 k \omega  y_5^{(0)}-2 \omega ^2 \left(35 y_1^{(0)}+47 y_2^{(0)}\right)\Big] +\nonumber\\
& & 2 \Big\lbrace k^2 \Big[(\Delta -3) \Delta  \Big(3 (16 (\Delta -3) \Delta+47) \lambda  v y_2^{(0)}+ 2 \mathcal{G}_2 \lambda  v y_4^{(0)}+72 (\Delta -1)(2 \Delta -1) v y_3^{(3-\Delta)}+\nonumber\\
& & 72 (\Delta -2) (2 \Delta -5) \lambda  y_3^{(\Delta)}\Big)-  36 \mathcal{G}_1 y_1^{(3)}-72 \mathcal{G}_1 y_4^{(3)}\Big]+4 k \omega \Big[(\Delta -3) \Delta  (\mathcal{G}_2) \lambda  v y_5^{(0)}- 36 \mathcal{G}_1 y_5^{(3)}\Big]+\nonumber\\
& & 2 \omega ^2 \Big[(\Delta -3) \Delta  \Big(-6 \left(8 \Delta^2-6 \Delta +1\right) v y_3^{(3-\Delta)}+(-16 (\Delta -3) \Delta -47) \lambda  v y_2^{(0)}-\nonumber\\
& & 6\left(8 \Delta ^2-42 \Delta +55\right) \lambda  y_3^{(\Delta)}\Big)-36 \mathcal{G}_1 y_1^{(3)}\Big]\Big\rbrace +2 (\Delta -3) \Delta  \mathcal{G}_2 \lambda  v y_1^{(0)}  \left(k^2+2 \omega ^2\right)\Bigg\rbrace \,,\nonumber\\
\mathcal{G}_1 &=& 4 (\Delta -3) \Delta +5,\,\,\mathcal{G}_2 = 8 (\Delta -3) \Delta -17,\nonumber\\
y_2^{(3\Delta)} &=& \frac{(3 \Delta +1) v}{128 \left(2 \Delta ^2+\Delta
		-1\right)} \Big\lbrace k^2 \left[v \left(y_1^{(0)}-3 y_2^{(0)}+2 y_4^{(0)}\right)-16
	y_3^{(\Delta)}\right]+4 k v \omega  y_5^{(0)}+\nonumber\\
	& & 2 \omega ^2 \left[v
	\left(y_1^{(0)}+y_2^{(0)}\right)+8 y_3^{(\Delta)}\right]\Big\rbrace\,, \nonumber\\
 y_2^{(4\Delta)} &=& -\frac{3 \lambda}{4 (2 \Delta -9) (4 \Delta -9)}  \Big\lbrace B_0 (9-4 \Delta ) \left[(2 \Delta -9)
 y_3^{(3-\Delta)}+(\Delta -3) \lambda 
 y_2^{(0)}\right]+\nonumber\\
 & & \Delta  (\Delta -3)^2 \lambda  \left(3 v
 y_3^{(3-\Delta)}+\lambda  v y_2^{(0)}+\lambda 
 y_3^{(\Delta)}\right)\Big\rbrace ,\,\nonumber
 \eea
 \begin{align}
 y_2^{(5\Delta)} =& \frac{9 \Delta  \lambda ^3 (4 y_3^{(3-\Delta)}+\lambda 
    y_2^{(0)})}{128 (4 \Delta -9)}, & y_4^{(1)} &= \frac {1}{4} \left[k^2 \left (y_1^{(0)} - y_2^{(0)} \right) + 2 k \omega  y_5^{(0)} + 2 \omega ^2 y_4^{(0)} \right],\,\nonumber\\
y_{4,5}^{(2)} &= -\frac{3}{16} \lambda ^2 y_{4,5}^{(0)}, & y_5^{(1)} &= \frac{1}{4} k \omega  \left(y_1^{(0)}+y_2^{(0)}-2y_4^{(0)}\right)\,.
\end{align}
For the auxiliary fields $\lbrace \eta_i\rbrace $, we write
\bea 
\e_{1,4,5} &=& e^{\frac{5}{2}r}\sum_{n>0}\left( \sum_{l=0}^4 e^{-l\Delta r}\eta_{1,4,5}^{(n,l)} + \sum_{l=1}^{4}e^{-l(3-\Delta) r}\eta_{1,4,5}^{(n,-l)}  \right) e^{-\frac{n}{2}r}\,, \nonumber\\
\e_2 &=& e^{3r}\sum_{n>0}\left( \e_2^{(n,-3)}e^{-3(2 - \Delta) r}+ \sum_{l=1,2,4} \e_2^{(n,-l)}e^{-l(3 - \Delta)r} + \sum_{l=0}^4 e^{-l\Delta r} \e_2^{(n,l)}   \right)e^{-\frac{n}{2} r}\,,\nonumber\\
\\
\e_3 &=& e^{\frac{3}{2}r}  \sum_{n>0} \left( \e_3^{(n,0)} +  \sum_{l=1}^{4} \e_3^{(n,-l)}e^{\Delta r}e^{-l(3 r - \Delta  r)} + \e_3^{(n,m,l)}e^{-(l\Delta -3)r}\right) e^{-\frac{n }{2}r}\,,\nonumber
\eea 
with (after imposing the boundary conditions \Eq{yetamap})
\bea 
\eta_1 &=& e^{\frac{5}{2}r} \Big\lbrace \eta_1^{(0,0)}  + y_1^{(0)} e^{-r} + \eta_1^{(4,0)} e^{-2r} +  \eta_1^{(0,-2)}e^{-2(3-\Delta)r} + \eta_1^{(6,0)} e^{-3r} + \eta_1^{(0,2)} e^{- 2\Delta r} +\nonumber\\
& &  \eta_1^{(2,-2)} e^{-(7-2\Delta)r} + y_1^{(3)} e^{-4r}+\cdots \Big\rbrace \,, \nonumber\\
\eta_2 &=& e^{3r}\Big\lbrace y_2^{(0)}  + \eta_2^{(4,0)} e^{-2r}  +\eta_2^{(0,-2)} e^{-2(3-\Delta)r}+\eta_2^{(6,0)}e^{-3r} + \eta_2^{(0,2)} e^{-2 \Delta  r} + \eta_2^{(8,0)} e^{-4r}  +\nonumber \\
& &  \eta_2^{(4,-2)} e^{-2(4-\Delta)r} + 
  \eta_2^{(10,0)} e^{-5 r} + \eta_2^{(4,2)} e^{-2(1+\Delta)r} + \eta_2^{(0,-4)} e^{-4(3-\Delta)r} +\nonumber\\
  & &  \eta_2^{(6,-2)} e^{-(9-2\Delta)r} + y_2^{(6)} e^{-6 r}  + \cdots \Big\rbrace\,, \\
\eta_3 &=& e^{\frac{3}{2}r}\Big\lbrace y_3^{(3-\Delta)} e^{-(3-\Delta) r} + y_3^{(\Delta)}e^{-\Delta r} +\eta_2^{(0)} \frac{\Delta-3}{2} \lambda e^{\Delta r}  +\cdots \Big\rbrace ,\nonumber\\
\eta_{4,5} &= &  e^{\frac{5}{2} r} \Big\lbrace \eta_{4,5}^{(0)} +  y_{4,5}^{(0)}e^{-r} +  \eta_{4,5}^{(4,0)}e^{-2r} + \eta_{4,5}^{(0,-2)} e^{-2(3-\Delta)r} +  \eta_{4,5}^{(6,0)} e^{-3r} + \eta_{4,5}^{(0,2)} e^{-2 \Delta  r} + \nonumber\\
& & \eta_{4,5}^{(2,-2)} e^{-(7-2\Delta)r} + y_{4,5}^{(3)} e^{-4r}  + \cdots \Big\rbrace \nonumber\,,
\eea

\begin{align}
\eta_1^{(4,0)} & =  -\frac{1}{16} y_2^{(0)} \left(k^4+k^2 \w^2-2 \w^4\right),& \eta_1^{(0)} &= -\frac{y_2^{(0)}}{8} \left(k^2+2 \w^2\right),\nonumber\\
\eta_1^{(0,-2)} &= \frac{(6 \Delta -13) \lambda ^2 y_2^{(0)} \left(k^2+2 \omega ^2\right)}{128 (2 \Delta -5)}, & \eta_1^{(0,2)} &= \frac{(6 \Delta -5) v^2 y_2^{(0)} \left(k^2+2 \omega ^2\right)}{128 (2 \Delta -1)}\,,\nonumber\\
\eta_1^{(2,-2)} &= -\frac{3}{16} \lambda ^2 y_1^{(0)},\,\, &\eta_2^{(4,0)} &= \frac{1}{24} y_2^{(0)}\left(2 w^2-5 k^2\right),\nonumber\\
\eta_2^{(0,-2)} &= \frac{\left(4 \Delta ^2-24 \Delta +45\right) }{8 (2 \Delta -9)}\lambda ^2 y_2^{(0)}, & \eta_2^{(8,0)} &= \frac{1}{48} y_2^{(0)} \left(k^4-5 k^2 \omega ^2+4 \omega^4\right), \nonumber\\
\eta_2^{(0,2)} &= -\frac{\left(4 \Delta ^2+9\right) }{8 (2\Delta +3)}v^2 y_2^{(0)}, & \eta_2^{(0,-4)} &= \frac{3888-\Delta  \big\lbrace 4 \Delta  \left[\Delta  (20 \Delta-141)+207\right]+1701\big\rbrace }{128 (2 \Delta-9) (2 \Delta -7) (4 \Delta -9)}\lambda ^4 y_2^{(0)} ,\nonumber\\
\eta_4^{(0,2)} &= \frac{(6 \Delta -5)}{64( 2\Delta -1)} k^2 v^2  y_2^{(0)}, & \eta_5^{(0,2)} &= \frac{(6 \Delta -5)}{32( 2\Delta -1)} k \omega  v^2 y_2^{(0)},\nonumber
\end{align}
\begin{align}
 \eta_4^{(4,0)} &= \frac{1}{8} k^2 y_2^{(0)} \left(\omega ^2-k^2\right)\,, &\eta_4^{(0,-2)} &= \frac{(6 \Delta -13)}{64 (2 \Delta-5)}k^2 \lambda ^2 y_2^{(0)},\nonumber\\
\eta_4^{(0)} &= -\frac{1}{4} k^2 y_2^{(0)}\nonumber\,, & \eta_4^{(2,-2)} &= -\frac{3}{16} \lambda ^2 y_4^{(0)}\,, & \eta_5^{(0)} &= -\frac{1}{2} k \omega  y_2^{(0)}\,,\nonumber\\
 \eta_5^{(4,0)} &= \frac{1}{4} k \omega  y_2^{(0)} \left(\omega ^2-k^2\right)\,, & \eta_5^{(0,-2)} &= \frac{(6 \Delta -13)}{32 (2\Delta -5)}k  \omega \lambda ^2 y_2^{(0)}\,, &\eta_5^{(2,-2)} &= -\frac{3}{16} \lambda ^2 y_5^{(0)}\,,\nonumber
\end{align}
\bea 
\eta_1^{(6,0)} & = & \frac{1}{144} \Bigg\lbrace \eta_2^{(0)} \left[B_0 \left(33 k^2-78 \omega ^2\right)-2 (\Delta -3)
\Delta  \lambda  v \left(k^2+2 \omega ^2\right)\right]+\nonumber\\
& & 12 \left[k^2 (3 y_4^{(0)}-4
y_1^{(0)}+3 k \omega  y_5^{(0)}-2 \omega ^2 y_2^{(0)} \right]\Bigg\rbrace ,\nonumber \\
\eta_2^{(4,-2)} &=& \frac{\lambda ^2 y_2^{(0)}}{192 \left(4
         \Delta ^3-44 \Delta ^2+157 \Delta -180\right)} \Big[  2 \left(48 \Delta ^4-600
         \Delta ^3+2830 \Delta ^2-5959 \Delta +4716\right) \omega
         ^2+\nonumber\\
& &\left(-96 \Delta ^4+1200 \Delta ^3-5654 \Delta
         ^2+11867 \Delta -9324\right) k^2\Big] ,\,\,\nonumber\\
\eta_2^{(6,0)} &=& \frac{3 y_2^{(0)} \left(B_0 (27-4 (\Delta -3) \Delta )-12
   (\Delta -3) \Delta  \lambda  v\right)}{3 (2 \Delta -9) (2 \Delta +3)}+ \frac{y_1^{(0)}}{3},\nonumber\\
\eta_2^{(10,0)} &=& \frac{1}{2160 (2
   \Delta -9) (2 \Delta +3)}\Big\lbrace k^2 \big[y_2^{(0)} \big[102 B_0 \left(4 \Delta
   ^2-12 \Delta -27\right)+\nonumber\\
& & \Delta  \left(532 \Delta ^3-3192
   \Delta ^2+14157 \Delta -28107\right) \lambda 
   v\big]-336 \left(4 \Delta ^2-12 \Delta -27\right)
   y_1^{(0)}+\nonumber\\
& & 36 \left(4 \Delta ^2-12 \Delta -27\right)
   y_4^{(0)}\big]+ 2 \omega ^2 \big\lbrace y_2^{(0)}
   \big[\Delta  \left(-236 \Delta ^3+1416 \Delta ^2-7011
   \Delta +14661\right) \lambda  v-\nonumber\\
& &   798 B_0 \left(4 \Delta
   ^2-12 \Delta -27\right)\big] -156 \left(4 \Delta ^2-12
   \Delta -27\right) y_1^{(0)}\big\rbrace +468 \left(4 \Delta
   ^2-12 \Delta -27\right) k \omega  y_5^{(0)}\Big\rbrace\,, \nonumber\\
\eta_2^{(4,2)} & =& \frac{v^2 y_2^{(0)}}{192 \left(4 \Delta ^3+8 \Delta
   ^2+\Delta -3\right)} \Big[ \left(96 \Delta ^4+48 \Delta
   ^3+38 \Delta ^2-25 \Delta -15\right) k^2-\nonumber\\
& & 2 \left(48
   \Delta ^4+24 \Delta ^3+22 \Delta ^2-5 \Delta -3\right)
   \omega ^2\Big]\,,\nonumber\\
\eta_2^{(6,-2)} &=& \frac{1}{96 (2 \Delta -15) (2
   \Delta -9) (2 \Delta -7) (2 \Delta +3) (4 \Delta -9)}\Big\lbrace (\Delta -3) \lambda  \nonumber\\
& &\big[-\lambda  y_2^{(0)}
   \big(24 B_0 (\Delta -3) (2 \Delta -7) (2 \Delta -3) (2
   \Delta +3) (4 \Delta -9)+\nonumber\\
& & \Delta  \lbrace 8 \Delta  [2 \Delta 
   (\Delta  \lbrace \Delta  [4 \Delta  (8 \Delta
   -133)+3669]-11889\rbrace +13500)+25839]-485757\rbrace \lambda 
   v\big)-\nonumber\\
& & 96 (2 \Delta -15) (2 \Delta -7) (2 \Delta +3) (4
   \Delta -9) y_3^{(3-\Delta)}\big]\Big\rbrace \,, \nonumber\\
\eta_4^{(6,0)} &=& \frac{1}{72} \left(k^2 \big\lbrace y_2^{(0)} \left[33 B_0-2(\Delta -3) \Delta  \lambda  v\right]+24 y_1^{(0)}\big\rbrace-36 k \omega  y_5^{(0)}+36 \omega ^2 y_4^{(0)}\right),\nonumber \\
\eta_5^{(6,0)} &= & -\frac{1}{36} k \omega  \Big\lbrace y_2^{(0)} \left[2 (\Delta -3)\Delta  \lambda  v-15 B_0\right]+12 y_1^{(0)}-18  y_4^{(0)}\Big\rbrace\,. \nonumber \\
\eea

\subsection{Fluctuations at the horizon}\label{app:flucexph}
	
From the indicial polynomials of the dynamical equations for each fluctuation, we infer $3$ roots,
\be
P_{0} = 0\,,\quad  P_\pm = \pm i\w c_H \,,\quad c_H = e^{-(A_H + B_H)}\,.
\ee

We demand regularity if the roots are real and ingoing condition if they are complex. Therefore, near the horizon, the original fluctuations may admit the general series expansion
\begin{align}
y_{i} &= \sum_{n=0} y^{(0,n)}_i(r-r_H)^{2n} + (r-r_H)^{P_-}\sum_{n=0} y^{(1,n)}_i(r-r_H)^{2n}\,,
\end{align}
with $i = 1\,,\cdots \,, 5 $ and we have ruled out the $P_+$ root. As stated in section \ref{sec:hocu}, the choice of the boundary conditions \Eq{yetamap} fixes the series expansions of the auxiliary fields, regardless if the $\lbrace y_i\rbrace $ fields have a well defined near-horizon behavior. Therefore, we shall consider
\bea
\eta_i &=& \sum_{n} \Big[\eta_i^{(0,n)}\left( r-r_H\right)^{n-2}  +
 (r-r_H)^{P_-}\eta^{(1,n)}_i (r-r_H)^{n} +\nonumber\\
 & &  (r-r_H)^{P_+}\eta^{(2,n)}_i (r-r_H)^{n} +  \eta_i^{(2,n)}\log \left(r-r_H\right)\Big]\,,\nonumber\\
\eea

The on-shell series expansions read
\bea
y_i &=& 
y_i^{(0,0)} + y_i^{(0,1)}
\left(r-r_H\right){}^2 +\left(r-r_H\right){}^{P_-}\left[ y_i^{(1,0)} +y_i^{(1,1)}
\left(r-r_H\right){}^2 \right]+\cdots\,,\quad i\neq 3,4\nonumber\\
y_3 &=& 
\left(r-r_H\right){}^{P_-}\left[ y_3^{(1,0)} +y_3^{(1,1)}
\left(r-r_H\right){}^2 \right]+\cdots\,,\nonumber\\
y_4 &=& 
 y_4^{(0,1)}
\left(r-r_H\right){}^2 +\left(r-r_H\right){}^{P_-}\left[ y_i^{(1,0)} +y_i^{(1,1)}
\left(r-r_H\right){}^2 \right]+\cdots\,,\nonumber\\
\eea
\bea
\eta_1 &=& \eta_1^{(0,2)} + \eta_1^{(1,0)}(r-r_H)^{P_-} + \eta_1^{(2,0)}(r-r_H)^{P_+}+ \eta_1^{(0,3)}(r-r_H) + \cdots\,,\nonumber\\
\eta_2 &=& \frac{\eta_2^{(0,1)}}{r-r_H} + \eta_2^{(0,2)} + \eta_2^{(1,0)} (r-r_H)^{P_-} + \eta_2^{(2,0)}(r-r_H)^{P_+} + \nonumber\\
& & (r-r_H)\left[ \eta_2^{(0,3)} + \eta_2^{(1,2)}(r-r_H)^{P_-} + \eta_2^{(2,2)}(r-r_H)^{P_+} \right] + \cdots\,,\nonumber\\
\eta_3 &=& \eta_3^{(1,0)}(r-r_H)^{P_-} + \eta_3^{(2,0)}(r-r_H)^{P_+} + \cdots \,,\nonumber\\
\eta_4 &=& \eta_4^{(0,2)} + \eta_4^{(1,0)}(r-r_H)^{P_-} + \eta_4^{(2,0)}(r-r_H)^{P_+} + \cdots \,,\nonumber\\
\eta_5 &=& \frac{\eta_5^{(0,0)}}{(r-r_H)^2} + \eta_5^{(0,2)}+ \eta_5^{(1,0)}(r-r_H)^{P_-}  + \eta_5^{(2,0)}(r-r_H)^{P_+} + 
 \cdots \,,\nonumber\\
\eea 
where most of the coefficients are not independent from each other. For the main fluctuations,
\bea
y_1{}^{\text{(0,1)}} &=& \frac{1}{6\left(\frac{4}{c_H^2} +\omega ^2\right)}\bigg\lbrace y_1{}^{\text{(0,0)}} \left[9 A_H{}^{\text{(1)}}
   \left(\frac{4}{c_H^2} +\omega ^2\right)-4 k^2 e^{2 B_H}\right]- 4 k\omega  y_5{}^{\text{(0,0)}} e^{2 B_H}\bigg\rbrace\,, \nonumber\\
y_1{}^{\text{(1,0)}} &= & \frac{i }{\omega }y_2{}^{\text{(1,0)}} e^{B_H-\frac{A_H}{2}}\,,	\nonumber\\
y_1{}^{\text{(1,1)}} &=& 
\frac{e^{B_H-\frac{3 A_H}{2}}}{4 \omega 
\left(\frac{1}{c_H}-i \omega \right) \left(
\frac{2}{c_H}-i \omega \right) \left(\frac{3}{c_H}-i \omega \right)} \Big\lbrace y_2{}^{\text{(1,0)}}
	\Big[ 2 e^{A_H} A_H^{\text{(1)}} \left(
	\frac{12 i \omega ^2}{c_H}+ \frac{21 \omega }{c_H^2} +\frac{18 i}{c_H^3} +11 \omega ^3\right)+\nonumber\\
& & k^2 e^{B_H{}} \left(\frac{1}{c_H}-3 i \omega \right)
	\left(\omega + \frac{2 i}{c_H}\right)\Big] -4 \omega 
	e^{\frac{3 A_H}{2}+B_H} \Big[k^2 y_4{}^{\text{(1,0)}} \left(
	\frac{2}{c_H}-i \omega \right)+\nonumber\\
& & y_3{}^{\text{(1,0)}}
	V'(\phi_H) e^{2 A_H{}} \left(
	\frac{3}{c_H}-2 i \omega \right)\Big]\Bigg\rbrace\,,\nonumber\\
y_2{}^{\text{(0,0)}} &=& -y_1{}^{\text{(0,0)}}e^{\frac{3 A_H}{2}}\,,\nonumber\\
y_2{}^{\text{(0,1)}} &=& -\frac{1}{3} e^{\frac{3 A_H}{2}} \left[9
   y_1{}^{\text{(0,0)}} A_H{}^{\text{(1)}} +\frac{4 k e^{2
   B_H} c_H \left(k y_1{}^{\text{(0,0)}}+\omega 
   y_5{}^{\text{(0,0)}}\right)}{\omega ^2 c_H+4}\right]\,,\nonumber\\
y_2{}^{\text{(1,1)}} &=& \frac{e^{B_H-A_H}}{4
\omega  \left(\frac{2}{c_H}-i \omega \right) \left(\frac{3}{c_H{}^2}-\frac{4 i \omega
}{c_H}-\omega ^2\right)} \Big\lbrace y_2{}^{\text{(1,0)}} \Bigg[-\frac{100 i \omega ^2}{c_H} A_H{}^{\text{(1)}} e^{2
A_H}+\frac{72 \omega}{c_H^2}  A_H^{\text{(1)}} e^{2
A_H}\nonumber\\
& & -36 \omega ^3 e^{2 A_H}
A_H^{\text{(1)}}- \frac{2 i k^2}{c_H^3}+\frac{2
	k^2 \omega }{c_H{}^2}-\frac{3 i k^2 \omega ^2}{c_H}-k^2 \omega ^3\Bigg]-\nonumber\\
& & \frac{4 k^2 \omega}{c_H^2} y_4{}^{\text{(1,0)}} e^{\frac{ A_H}{2}}-4 \omega 
y_3{}^{\text{(1,0)}} V'(\phi_H) e^{\frac{7 A_H}{2}}
\left(\frac{3}{c_H{}^2}-\frac{3 i \omega }{c_H}-\omega ^2\right) \Bigg\rbrace\,,\nonumber\\
y_3{}^{\text{(1,1)}} &=&  \frac{1}{8 \left(\frac{1}{c_H}-i \omega
	\right)}\Bigg[2 y_3{}^{\text{(1,0)}} \left(k^2
	e^{B_H{}-A_H{}}+\frac{6 A_H{}^{\text{(1)}}}{c_H}-2 i \omega 
	A_H{}^{\text{(1)}}+\frac{V''(\phi_H)}{c_H}\right)+\nonumber\\
	& & i \omega  y_2{}^{\text{(1,0)}}
	V'(\phi_H) e^{-\frac{3 A_H{}}{2}}\Bigg]\,,\nonumber\\
y_4^{(0,1)} &=& -\frac{k e^{2 B_H} \left(k
   y_1{}^{\text{(0,0)}}+\omega 
   y_5{}^{\text{(0,0)}}\right)}{\frac{4}{c_H^2}+\omega ^2}\,,\nonumber\\
y_4{}^{\text{(1,1)}} &=& \frac{1}{\frac{8
	\omega }{c_H}-8 i \omega ^2}\Bigg[ y_2{}^{\text{(1,0)}}k^2 e^{B_H{}-\frac{3
			A_H{}}{2}} \left(- \omega  e^{-A_H{}}+i e^{ B_H{}}\right) + 2 k \omega ^2 y_5{}^{\text{(1,0)}}
e^{B_H{}-A_H{}}+\nonumber\\
& & y_4{}^{\text{(1,0)}} \left(\frac{12 \omega A_H{}^{\text{(1)}}}{c_H} - 4 i \omega ^2 A_H{}^{\text{(1)}}\right)
\Bigg]\,, \nonumber\\
y_5^{(0,1)} &=& \frac{3}{2} y_5{}^{\text{(0,0)}} A_H{}^{\text{(1)}}\,,\nonumber\\
y_5{}^{\text{(1,0)}} &= &\frac{k}{2 \omega  \left(\omega +\frac{2i}{c_H}\right)}\left[ y_2{}^{\text{(1,0)}} e^{-\frac{A_H}{2}}\left(\omega  e^{-A_H} + ie^{B_H}\right) - 2 \w y_4{}^{\text{(1,0)}}\right]\,,\nonumber\\
\eea
whilst for the auxiliary fluctuations,
\bea
\eta_1^{(0,2)} &=& -\frac{k \tilde{\eta}^H_5 K_{55}^{\text{H}}+2
   \omega  e^{\frac{3 A_H}{2}} \tilde{\eta}^H_2
   K_{22}^{\text{H}}}{2 \omega  K_{11}^{\text{H}}}\nonumber\,,\\
\eta_1^{(0,3)} &=& \frac{\omega  e^{\frac{3 A_H}{2}}}{2
   \left(\frac{1}{c_H^2}+\omega ^2\right)
   \left(K_{11}^{\text{H}}\right){}^2} \left[k \tilde{\eta}^H
   _5 K_{12}^{\text{H}}
   K_{55}^{\text{H}}+2 \omega  K_{22}^{\text{H}}
   \left(e^{\frac{3 A_H}{2}} \tilde{\eta}_2^H
   K_{12}^{\text{H}}-\eta_2^H
   K_{11}^{\text{H}}\right)\right]\,,\nonumber\\
\eta_2^{(1,0)} &=& -\frac{\eta_1^H K_{12}^{\text{H}}+\eta_3^H
   K_{32}^{\text{H}}}{K_{22}^{\text{H}}}\,,\nonumber\\
\eta_2^{(2,0)} &=& -\frac{\tilde{\eta}_1^H K_{12}^{\text{H}}+\tilde{\eta}_3^H K_{32}^{\text{H}}}{K_{22}^{\text{H}}}\,,\nonumber\\
\eta_2^{(0,3)} &=& \frac{e^{-\frac{3 A_H}{2}}}{2 \omega  K_{22}^{\text{H}}} \Bigg\lbrace A_H{}^{\text{(1)}} \left(8
   \omega  e^{\frac{3 A_H}{2}} \tilde{\eta}_2^H
   K_{22}^{\text{H}}-3 k \tilde{\eta}_5^H
   K_{55}^{\text{H}}\right)-\nonumber\\
& & \frac{\omega ^2 e^{3 A_H}
   K_{12}^{\text{H}} \left[K_{12}^{\text{H}} \left(k
   \tilde{\eta}^H_5 K_{55}^{\text{H}}+2 \omega 
   e^{\frac{3 A_H}{2}} \tilde{\eta}^H_2
   K_{22}^{\text{H}}\right)-2 \omega  \eta^H_2 K_{11}{}^{\text{H}}
   K_{22}^{\text{H}}\right]}{\left(K_{11}^{\text{H}}\right){}^2 \left(\frac{1}{c_H^2}+\omega
   ^2\right)}\Bigg\rbrace \,,\nonumber\\
\eta_2^{(1,2)} & = & \frac{e^{-\frac{3 A_H}{2}}}{2 K^H_{22}{}
   \left( -\frac{3 i \omega}{c_H}+\frac{2}{c_H^2}-\omega ^2\right)} \Big\lbrace 2
   \eta^H_1 K^H_{11}{} \Big[2 k^2 e^{2
   B_H{}}+ A_H{}^{\text{(1)}}  \left( 
   \frac{7 i \omega}{c_H}+\frac{6}{c_H^2} +5
   \omega ^2\right)\Big]+\nonumber\\
& & \eta^H_3 \omega  K^H_{33} V'(\phi_H) \left(\omega + \frac{2 i}{c_H}\right)+
 2k^2 \eta^H_4 K^H_{44}{} e^{2B_H{}}\Big\rbrace ,\nonumber\\
\eta_2^{(2,2)} &=& \frac{e^{-\frac{3 A_H}{2}}}{2
   K_{22}{}^{\text{H}} \left(3 i \frac{\omega}{c_H}  +\frac{2}{c_H^2} -\omega ^2\right)} \Bigg\lbrace 2 \tilde{\eta}_1^H
   K_{11}{}^{\text{H}} \left[2 k^2 e^{2
   B_H}+A_H{}^{\text{(1)}} \left(-7 i \frac{\omega}{c_H}  +\frac{6}{c_H^2} +5 \omega
   ^2\right)\right]+\nonumber\\
   & & \omega  \tilde{\eta}_3^H
   K_{33}{}^{\text{H}} V'(\phi_H) \left(\omega-2 i e^{A_H+B_H}\right)+2 k^2 e^{2 B_H} \tilde{\eta}^H_4 K_{44}{}^{\text{H}}\Bigg\rbrace \,,\nonumber\\
 \eta_4^{(0,2)} &=& \frac{k \tilde{\eta}^H_5 K_{55}^{\text{H}}}{\omega K_{44}^{\text{H}}}\,,\nonumber\\ 
 \eta_5^{(1,0)} &=& -\frac{i k e^{2 B_H} \left(2 \eta^H_1
    K_{11}^{\text{H}}+\eta^H_4
    K_{44}^{\text{H}}\right)}{K_{55}^{\text{H}} \left(\frac{2}{c_H}-i \omega \right)}\,,\nonumber\\
\eta_5^{(2,0)} &=& \frac{i k e^{2 B_H} \left(2 \tilde{\eta}^H_1
   K_{11}^{\text{H}}+\tilde{\eta}^H_4
   K_{44}^{\text{H}}\right)}{\left(\frac{2}{c_H}+i \omega
   \right) K_{55}^{\text{H}}}\,,\nonumber\\
\eea
plus higher order terms. $\eta_{1,3,4}^{(1,0)} = \eta_{1,3,4}^H\,,\eta_{1,3,4}^{(2,0)} = \tilde{\eta}_{1,3,4}^H\,, \eta^{(0,2)}_2 = \eta_2^H\,, \eta_2^{(0,1)} = \tilde{\eta}_2^H\,, \eta_5^{(0,2)} = \eta_5^H$ and $\eta_5^{(0,0)} = \tilde{\eta_5}^H$, as it appears in section \ref{sec:hocu}.\\

\bibliographystyle{JHEP}
\bibliography{holoWard}

\providecommand{\href}[2]{#2}\begingroup\raggedright\begin{thebibliography}{10}

\bibitem{Maldacena:1997re}
J.~M. Maldacena, {\it {The large N limit of superconformal field theories and
  supergravity}},  {\em Adv. Theor. Math. Phys.} {\bf 2} (1998) 231--252,
  [\href{http://arxiv.org/abs/9711200}{{\tt 9711200}}].

\bibitem{Gubser:1998bc}
S.~S. Gubser, I.~R. Klebanov, and A.~M. Polyakov, {\it {Gauge theory
  correlators from noncritical string theory}},  {\em Phys.Lett.} {\bf B428}
  (1998) 105--114, [\href{http://arxiv.org/abs/hep-th/9802109}{{\tt
  hep-th/9802109}}].

\bibitem{Witten:1998qj}
E.~Witten, {\it {Anti-de Sitter space and holography}},  {\em
  Adv.Theor.Math.Phys.} {\bf 2} (1998) 253--291,
  [\href{http://arxiv.org/abs/hep-th/9802150}{{\tt hep-th/9802150}}].

\bibitem{Policastro:2001yc}
G.~Policastro, D.~T. Son, and A.~O. Starinets, {\it {The Shear viscosity of
  strongly coupled N=4 supersymmetric Yang-Mills plasma}},  {\em Phys. Rev.
  Lett.} {\bf 87} (2001) 081601,
  [\href{http://arxiv.org/abs/hep-th/0104066}{{\tt hep-th/0104066}}].

\bibitem{Bhattacharyya:2008jc}
S.~Bhattacharyya, V.~E. Hubeny, S.~Minwalla, and M.~Rangamani, {\it {Nonlinear
  Fluid Dynamics from Gravity}},  {\em JHEP} {\bf 02} (2008) 045,
  [\href{http://arxiv.org/abs/0712.2456}{{\tt arXiv:0712.2456}}].

\bibitem{Banerjee:2012iz}
N.~Banerjee, J.~Bhattacharya, S.~Bhattacharyya, S.~Jain, S.~Minwalla, and
  T.~Sharma, {\it {Constraints on Fluid Dynamics from Equilibrium Partition
  Functions}},  {\em JHEP} {\bf 09} (2012) 046,
  [\href{http://arxiv.org/abs/1203.3544}{{\tt arXiv:1203.3544}}].

\bibitem{Jensen:2012jh}
K.~Jensen, M.~Kaminski, P.~Kovtun, R.~Meyer, A.~Ritz, and A.~Yarom, {\it
  {Towards hydrodynamics without an entropy current}},  {\em Phys. Rev. Lett.}
  {\bf 109} (2012) 101601, [\href{http://arxiv.org/abs/1203.3556}{{\tt
  arXiv:1203.3556}}].

\bibitem{Hoyos:2011ez}
C.~Hoyos and D.~T. Son, {\it {Hall Viscosity and Electromagnetic Response}},
  {\em Phys. Rev. Lett.} {\bf 108} (2012) 066805,
  [\href{http://arxiv.org/abs/1109.2651}{{\tt arXiv:1109.2651}}].

\bibitem{PhysRevLett.69.953}
X.~G. Wen and A.~Zee, {\it Shift and spin vector: New topological quantum
  numbers for the hall fluids},  {\em Phys. Rev. Lett.} {\bf 69} (Aug, 1992)
  953--956.

\bibitem{PhysRevB.79.045308}
N.~Read, {\it Non-abelian adiabatic statistics and hall viscosity in quantum
  hall states and ${p}_{x}+i{p}_{y}$ paired superfluids},  {\em Phys. Rev. B}
  {\bf 79} (Jan, 2009) 045308.

\bibitem{PhysRevB.84.085316}
N.~Read and E.~H. Rezayi, {\it Hall viscosity, orbital spin, and geometry:
  Paired superfluids and quantum hall systems},  {\em Phys. Rev. B} {\bf 84}
  (Aug, 2011) 085316.

\bibitem{Bradlyn:2012ea}
B.~Bradlyn, M.~Goldstein, and N.~Read, {\it {Kubo formulas for viscosity: Hall
  viscosity, Ward identities, and the relation with conductivity}},  {\em Phys.
  Rev.} {\bf B86} (2012) 245309, [\href{http://arxiv.org/abs/1207.7021}{{\tt
  arXiv:1207.7021}}].

\bibitem{Hoyos:2015yna}
C.~Hoyos, B.~S. Kim, and Y.~Oz, {\it {Ward Identities for Transport in 2+1
  Dimensions}},  {\em JHEP} {\bf 1503} (2015) 164,
  [\href{http://arxiv.org/abs/1501.05756}{{\tt arXiv:1501.05756}}].

\bibitem{deHaro:2000xn}
S.~de~Haro, S.~N. Solodukhin, and K.~Skenderis, {\it {Holographic
  reconstruction of space-time and renormalization in the AdS / CFT
  correspondence}},  {\em Commun. Math. Phys.} {\bf 217} (2001) 595--622,
  [\href{http://arxiv.org/abs/hep-th/0002230}{{\tt hep-th/0002230}}].

\bibitem{Papadimitriou:2004rz}
I.~Papadimitriou and K.~Skenderis, {\it {Correlation functions in holographic
  RG flows}},  {\em JHEP} {\bf 10} (2004) 075,
  [\href{http://arxiv.org/abs/hep-th/0407071}{{\tt hep-th/0407071}}].

\bibitem{Hartnoll:2007ip}
S.~A. Hartnoll and C.~P. Herzog, {\it {Ohm's Law at strong coupling: S duality
  and the cyclotron resonance}},  {\em Phys. Rev.} {\bf D76} (2007) 106012,
  [\href{http://arxiv.org/abs/0706.3228}{{\tt arXiv:0706.3228}}].

\bibitem{Herzog:2009xv}
C.~P. Herzog, {\it {Lectures on Holographic Superfluidity and
  Superconductivity}},  {\em J. Phys.} {\bf A42} (2009) 343001,
  [\href{http://arxiv.org/abs/0904.1975}{{\tt arXiv:0904.1975}}].

\bibitem{Lindgren:2015lia}
J.~Lindgren, I.~Papadimitriou, A.~Taliotis, and J.~Vanhoof, {\it {Holographic
  Hall conductivities from dyonic backgrounds}},  {\em JHEP} {\bf 07} (2015)
  094, [\href{http://arxiv.org/abs/1505.04131}{{\tt arXiv:1505.04131}}].

\bibitem{Kanitscheider:2008kd}
I.~Kanitscheider, K.~Skenderis, and M.~Taylor, {\it {Precision holography for
  non-conformal branes}},  {\em JHEP} {\bf 09} (2008) 094,
  [\href{http://arxiv.org/abs/0807.3324}{{\tt arXiv:0807.3324}}].

\bibitem{Papadimitriou:2011qb}
I.~Papadimitriou, {\it {Holographic Renormalization of general dilaton-axion
  gravity}},  {\em JHEP} {\bf 08} (2011) 119,
  [\href{http://arxiv.org/abs/1106.4826}{{\tt arXiv:1106.4826}}].

\bibitem{Gouteraux:2011qh}
B.~Gouteraux, J.~Smolic, M.~Smolic, K.~Skenderis, and M.~Taylor, {\it
  {Holography for Einstein-Maxwell-dilaton theories from generalized
  dimensional reduction}},  {\em JHEP} {\bf 01} (2012) 089,
  [\href{http://arxiv.org/abs/1110.2320}{{\tt arXiv:1110.2320}}].

\bibitem{Caldarelli:2013aaa}
M.~M. Caldarelli, J.~Camps, B.~Goutéraux, and K.~Skenderis, {\it
  {AdS/Ricci-flat correspondence}},  {\em JHEP} {\bf 04} (2014) 071,
  [\href{http://arxiv.org/abs/1312.7874}{{\tt arXiv:1312.7874}}].

\bibitem{Chemissany:2014xsa}
W.~Chemissany and I.~Papadimitriou, {\it {Lifshitz holography: The whole
  shebang}},  {\em JHEP} {\bf 01} (2015) 052,
  [\href{http://arxiv.org/abs/1408.0795}{{\tt arXiv:1408.0795}}].

\bibitem{Hartong:2014oma}
J.~Hartong, E.~Kiritsis, and N.~A. Obers, {\it {Lifshitz space–times for
  Schrödinger holography}},  {\em Phys. Lett.} {\bf B746} (2015) 318--324,
  [\href{http://arxiv.org/abs/1409.1519}{{\tt arXiv:1409.1519}}].

\bibitem{Son:2002sd}
D.~T. Son and A.~O. Starinets, {\it {Minkowski space correlators in AdS / CFT
  correspondence: Recipe and applications}},  {\em JHEP} {\bf 09} (2002) 042,
  [\href{http://arxiv.org/abs/hep-th/0205051}{{\tt hep-th/0205051}}].

\bibitem{Amado:2009ts}
I.~Amado, M.~Kaminski, and K.~Landsteiner, {\it {Hydrodynamics of Holographic
  Superconductors}},  {\em JHEP} {\bf 05} (2009) 021,
  [\href{http://arxiv.org/abs/0903.2209}{{\tt arXiv:0903.2209}}].

\bibitem{Lee:1990:LSC}
J.~Lee and R.~M. Wald, {\it Local symmetries and constraints},  {\em J. of
  Math. Phys.} {\bf 31} (Mar., 1990) 725--743.

\bibitem{Wald:1999wa}
R.~M. Wald and A.~Zoupas, {\it {A General definition of 'conserved quantities'
  in general relativity and other theories of gravity}},  {\em Phys. Rev.} {\bf
  D61} (2000) 084027, [\href{http://arxiv.org/abs/gr-qc/9911095}{{\tt
  gr-qc/9911095}}].

\bibitem{Papadimitriou:2005ii}
I.~Papadimitriou and K.~Skenderis, {\it {Thermodynamics of asymptotically
  locally AdS spacetimes}},  {\em JHEP} {\bf 08} (2005) 004,
  [\href{http://arxiv.org/abs/hep-th/0505190}{{\tt hep-th/0505190}}].

\bibitem{Papadimitriou:2010as}
I.~Papadimitriou, {\it {Holographic renormalization as a canonical
  transformation}},  {\em JHEP} {\bf 11} (2010) 014,
  [\href{http://arxiv.org/abs/1007.4592}{{\tt arXiv:1007.4592}}].

\bibitem{Hartnoll:2007ai}
S.~A. Hartnoll and P.~Kovtun, {\it {Hall conductivity from dyonic black
  holes}},  {\em Phys. Rev.} {\bf D76} (2007) 066001,
  [\href{http://arxiv.org/abs/0704.1160}{{\tt arXiv:0704.1160}}].

\bibitem{Hartnoll:2007ih}
S.~A. Hartnoll, P.~K. Kovtun, M.~Muller, and S.~Sachdev, {\it {Theory of the
  Nernst effect near quantum phase transitions in condensed matter, and in
  dyonic black holes}},  {\em Phys. Rev.} {\bf B76} (2007) 144502,
  [\href{http://arxiv.org/abs/0706.3215}{{\tt arXiv:0706.3215}}].

\bibitem{Goldstein:2010aw}
K.~Goldstein, N.~Iizuka, S.~Kachru, S.~Prakash, S.~P. Trivedi, and A.~Westphal,
  {\it {Holography of Dyonic Dilaton Black Branes}},  {\em JHEP} {\bf 10}
  (2010) 027, [\href{http://arxiv.org/abs/1007.2490}{{\tt arXiv:1007.2490}}].

\bibitem{Gubankova:2010rc}
E.~Gubankova, J.~Brill, M.~Cubrovic, K.~Schalm, P.~Schijven, and J.~Zaanen,
  {\it {Holographic fermions in external magnetic fields}},  {\em Phys. Rev.}
  {\bf D84} (2011) 106003, [\href{http://arxiv.org/abs/1011.4051}{{\tt
  arXiv:1011.4051}}].

\bibitem{Lippert:2014jma}
M.~Lippert, R.~Meyer, and A.~Taliotis, {\it {A holographic model for the
  fractional quantum Hall effect}},  {\em JHEP} {\bf 01} (2015) 023,
  [\href{http://arxiv.org/abs/1409.1369}{{\tt arXiv:1409.1369}}].

\bibitem{Fujita:2009kw}
M.~Fujita, W.~Li, S.~Ryu, and T.~Takayanagi, {\it {Fractional Quantum Hall
  Effect via Holography: Chern-Simons, Edge States, and Hierarchy}},  {\em
  JHEP} {\bf 06} (2009) 066, [\href{http://arxiv.org/abs/0901.0924}{{\tt
  arXiv:0901.0924}}].

\bibitem{Bergman:2010gm}
O.~Bergman, N.~Jokela, G.~Lifschytz, and M.~Lippert, {\it {Quantum Hall Effect
  in a Holographic Model}},  {\em JHEP} {\bf 10} (2010) 063,
  [\href{http://arxiv.org/abs/1003.4965}{{\tt arXiv:1003.4965}}].

\bibitem{Jokela:2011eb}
N.~Jokela, M.~Jarvinen, and M.~Lippert, {\it {A holographic quantum Hall model
  at integer filling}},  {\em JHEP} {\bf 05} (2011) 101,
  [\href{http://arxiv.org/abs/1101.3329}{{\tt arXiv:1101.3329}}].

\bibitem{Kristjansen:2012ny}
C.~Kristjansen and G.~W. Semenoff, {\it {Giant D5 Brane Holographic Hall
  State}},  {\em JHEP} {\bf 06} (2013) 048,
  [\href{http://arxiv.org/abs/1212.5609}{{\tt arXiv:1212.5609}}].

\bibitem{Bea:2014yda}
Y.~Bea, N.~Jokela, M.~Lippert, A.~V. Ramallo, and D.~Zoakos, {\it {Flux and
  Hall states in ABJM with dynamical flavors}},  {\em JHEP} {\bf 03} (2015)
  009, [\href{http://arxiv.org/abs/1411.3335}{{\tt arXiv:1411.3335}}].

\bibitem{KeskiVakkuri:2008eb}
E.~Keski-Vakkuri and P.~Kraus, {\it {Quantum Hall Effect in AdS/CFT}},  {\em
  JHEP} {\bf 09} (2008) 130, [\href{http://arxiv.org/abs/0805.4643}{{\tt
  arXiv:0805.4643}}].

\bibitem{Fujita:2012fp}
M.~Fujita, M.~Kaminski, and A.~Karch, {\it {SL(2,Z) Action on AdS/BCFT and Hall
  Conductivities}},  {\em JHEP} {\bf 07} (2012) 150,
  [\href{http://arxiv.org/abs/1204.0012}{{\tt arXiv:1204.0012}}].

\bibitem{Saremi:2011ab}
O.~Saremi and D.~T. Son, {\it {Hall viscosity from gauge/gravity duality}},
  {\em JHEP} {\bf 04} (2012) 091, [\href{http://arxiv.org/abs/1103.4851}{{\tt
  arXiv:1103.4851}}].

\bibitem{Son:2013xra}
D.~T. Son and C.~Wu, {\it {Holographic Spontaneous Parity Breaking and Emergent
  Hall Viscosity and Angular Momentum}},  {\em JHEP} {\bf 07} (2014) 076,
  [\href{http://arxiv.org/abs/1311.4882}{{\tt arXiv:1311.4882}}].

\bibitem{Hoyos:2014nua}
C.~Hoyos, B.~S. Kim, and Y.~Oz, {\it {Odd Parity Transport In Non-Abelian
  Superfluids From Symmetry Locking}},  {\em JHEP} {\bf 10} (2014) 127,
  [\href{http://arxiv.org/abs/1404.7507}{{\tt arXiv:1404.7507}}].

\bibitem{Golkar:2015dya}
S.~Golkar and M.~M. Roberts, {\it {Viscosities and shift in a chiral
  superfluid: a holographic study}},
  \href{http://arxiv.org/abs/1502.07690}{{\tt arXiv:1502.07690}}.

\end{thebibliography}\endgroup

\end{document}